\documentclass[aps,showpacs,preprintnumbers,amsmath,nofootinbib]{revtex4}
\usepackage{amssymb}
\usepackage{graphicx}
\usepackage{dcolumn}
\usepackage{bm}
\usepackage{color}
\usepackage{footnpag}

\begin{document}


\newcommand{\blue}[1]{\textcolor{blue}{#1}}
\newcommand{\new}{\blue}
\newcommand{\green}[1]{\textcolor{green}{#1}}
\newcommand{\modif}{\green}

\title{Phenomenological Analysis of $pp$ and $\bar{p}p$ Elastic Scattering Based on Theoretical
Bounds in High-Energy Physics}

\author{S. D. Campos}
\email{sergiodc@ufscar.br}
\affiliation{Universidade Federal de S\~ao Carlos, campus de Sorocaba, 18052-780, Sorocaba, SP, Brazil.}

\author{V. A. Okorokov}
\email{okorokov@bnl.gov; VAOkorokov@mephi.ru}
\affiliation{National Research Nuclear University "MEPhI",
Kashirskoe Shosse 31, 115409 Moscow, Russia}


\begin{abstract}

Considering the Froissart-Martin bound, Jin-Martin-Cornille bound and the optical theorem, we propose a novel
parametrization for the total cross-section of proton-proton and antiproton-proton elastic scattering data. Using derivative
dispersion relations we obtain the real part of the elastic scattering amplitude and thus the $\rho$ parameter. Simultaneous
fits to $\sigma_{\footnotesize\mbox{tot}}$ and $\rho$ are performed allowing very good statistical descriptions of the
available data. Furthermore, predictions to $\sigma_{\footnotesize\mbox{tot}}$ and $\rho$ at energies not used in the fit procedures are presented. For $\sigma_{\footnotesize\mbox{tot}}$ we obtain predictions at RHIC, LHC and future HC energies.

\end{abstract}

\pacs{13.85.Dz; 13.85.-t}

\maketitle

\section{\label{intro}Introduction}
Nowadays, the study of high energy scattering is one of the most exciting topics in physics, and we have, as far we know, the exact
theory of strong interactions, Quantum Chromodynamics (QCD), which can describe, based on the perturbative calculation scheme, the
hadron-hadron interactions at short distances. However, interactions at large distances (near forward scattering) cannot
be calculated even within perturbative approaches. Proton-proton ($pp$) and antiproton-proton ($\bar{p}p$) elastic scattering
are the simplest processes not yet explained in terms of a pure QCD description. The absence of such a description allows
empirical analyses based on models
\cite{modelo1,modelo2,modelo3,modelo4,modelo5,modelo6,modelo7,modelo8,modelo9,modelo10,modelo11,modelo12,modelo13,modelo14}
or almost-independent models (some experimental data set possess theoretical bias) \cite{sdc} whose main goal is to fit the available experimental data extracting useful information, contributing to the development of novel computational schemes. Several phenomenological models for high energy scattering are
based on general principles of the Axiomatic Quantum Field Theory (AQFT), such as unitarity and crossing-symmetry, and on derived results as analyticity, and have proven to be successful in understanding or predicting the behavior of the hadronic scattering amplitude.

Empirical parametrization for the total cross-section, differential cross-section, slope of the differential
cross-section and others physical quantities have widely been used as a source of phenomenological model-independent indications of the experimental data behavior. These approaches must be seen as a first step or attempt toward a formally rigorous phenomenological model-independent description of high energy elastic hadron scattering, embodying a predictive character. Moreover, important high energy theorems and bounds have been demonstrated in the last decades, providing rigorous formal constraints in the region of asymptotic energies, which cannot be disregarded in any reliable formalism, mainly related with phenomenological model-independent approaches. Of course, we must keep in mind that AQFT is a mathematical framework for the treatment and interpretation of relativistic quantum field theories and, in general, not all formal results based on AQFT results are satisfied by QCD.

In this work, we present a novel parametrization based on formal high energy bounds taken from AQFT for the total cross-section of
the elastic scattering amplitude. Specifically, we introduce a novel analytical parametrization for the total cross-section,
inferred assuming some rigorous high energy bounds. The real part of the scattering amplitude is analytically evaluated through
derivative dispersion relation. The crossing symmetry relates particle-particle and particle-antiparticle amplitudes. Fits to
$pp$ and $\bar{p}p$ total cross-section, $\rho$ parameter are performed and may lead to a predictive and phenomenological model-independent approach, able to describe quite well the bulk of the experimental data. Furthermore, we present some predictions for the total cross-section for the LHC run at $\sqrt{s}=7.0$, 10, and 14 TeV,
for the future HC at $\sqrt{s}=50$, 100, 200 and 500 TeV and for the RHIC experiment at $\sqrt{s}=200$ and 500 GeV. It is important
to stress that the parametrization proposed here allows us to obtain, under some physical conditions, the whole scattering
function (beyond the forward direction). We intend to show these analyses elsewhere.

The paper is organized as follows. In section \ref{sec2}, the Froissart-Martin bound \cite{froissartmartin1,froissartmartin2}, Jin-Martin-Cornille bound \cite{jinmartin,cornille} and the optical theorem are used to propose a novel parameterization for $pp$ and $\bar{p}p$ total cross-section and $\rho$ parameter using only four fit parameters to each scattering reaction. In section \ref{sec4} we explicitly obtain the real part of the forward elastic scattering amplitude and the $\rho$ parameter. In section \ref{sec3} we show the individual fit results for $\sigma_{\footnotesize\mbox{tot}}$ and $\rho$. In section \ref{secSimult} we show the simultaneous fit results for $\sigma_{\footnotesize\mbox{tot}}$ and $\rho$. Section \ref{secY} presents a possible interpretation for the effective physical mechanism of this model comparing it with some models present in the literature. In section \ref{sec7} we present our Final Remarks with some criticism.

\section{\label{sec2}Total Cross-Section}
The total cross-section is one of the most important physical observable in the elastic scattering. Its energy dependence has
been a focus of intense theoretical interest since the establishment of QCD as $the$ theory of strong interactions.
However, at high energies there is no description based only on pure QCD formalism and the use of models with free fit parameters
is the usual way to study $pp$ and $\bar{p}p$ elastic scattering. In this section one obtains a novel parametrization for the total
cross-section based on well-established results taken from AQFT.

We start considering the scattering amplitude $F(s,q^2)$ expressed as a function of two Mandelstam variables in the center-of-mass
system, usually the energy squared $s$ and the momentum transfer squared $t=-q^2$. We can write $F(s,q^2)$ in terms of its real and
imaginary parts as
\begin{eqnarray}
\nonumber F(s,q^2)=\mathrm{Re}F(s,q^2)+i\mathrm{Im}F(s,q^2),
\end{eqnarray}

\noindent where the imaginary part of $F$ represents the absorption in the scattering process and usually is related to the
real part by dispersion relation (integral or derivative) that analytically connects $pp$ and $\bar{p}p$ elastic scattering
through crossing property. Therefore, if we know $\mathrm{Im}F(s,q^2)$, then we are able to compute analytically
the whole scattering amplitude (at $q^2=0$).

The main goal of this section is to propose $pp$ and $\bar{p}p$ total cross-section considering high energy bounds from formal results of the AQFT. A fundamental result derived from unitarity and analyticity asserts that high energy total cross-sections for the hadronic forward elastic scattering (at $q^2=0$) should be bounded according to Froissart-Martin bound \cite{froissartmartin1,froissartmartin2}
\begin{eqnarray}
\label{1}\sigma_{\footnotesize\mbox{tot}}\leq
\frac{\pi}{m_{\pi}^2}\ln^2(s/s_0),
\end{eqnarray}

\noindent where $\sigma_{\footnotesize\mbox{tot}}=\sigma_{\footnotesize\mbox{tot}}(s)$
is the total cross-section of $pp$ and $\bar{p}p$ and $s_0$ is some energy initial value. Henceforth, we shall adopt $s_0=1$ GeV$^2$. This bound refers to an energy dependence of the total cross-section rising no more rapidly than $\ln^2s$. As a remark, unfortunately, at present day we cannot be unambiguously discriminate between asymptotic fits of $\ln s$ and $\ln^2 s$ using high energy data \cite{block,compete1,compete2}.

On the other hand, Cornille \cite{cornille} improving the result of Jin and Martin \cite{jinmartin} proved the asymptotic lower bound
\begin{eqnarray}
\label{2} \sigma_{\footnotesize\mbox{tot}} \geq \frac{\delta}{s^6}.
\end{eqnarray}

Using both formal results (\ref{1}) and (\ref{2}) we propose the following parametrization for the total cross-section

\begin{eqnarray}
\label{pp} \sigma_{\footnotesize\mbox{tot}}^{pp} =
\frac{\delta}{s^{6+\gamma}} +
\frac{\pi\beta}{\sqrt{2}s^{\alpha}m_{\pi}^2} \ln^2s,
\end{eqnarray}
\vspace*{-0.5cm}
\begin{eqnarray}
\label{pbp} \sigma_{\footnotesize\mbox{tot}}^{\bar{p}p} =
\frac{\bar{\delta}}{s^{6+\bar{\gamma}}} + \frac{\pi\bar{\beta}}
{\sqrt{2}s^{\bar{\alpha}}m_{\pi}^2} \ln^2s.
\end{eqnarray}

The first term in the right hand side of (\ref{pp}) and (\ref{pbp}) will be responsible for the description of the data at
low energies. Therefore, $\delta$ and $\bar{\delta}$, $\gamma$ and $\bar{\gamma}$ will be slightly different and at very high
energies both terms must vanish. Yet, $\alpha$ and $\bar{\alpha}$ also has a role at low energies allowing a more accurate
description of the low energy data. Nevertheless, the second term in the right hand side of both parameterizations will be the
leading one at high energies. Using the Pomeranchuk theorem, $\beta$ and $\bar{\beta}$, $\alpha$ and $\bar{\alpha}$ must tend
to the same numerical value.

As seen above, in the fitting process for the total cross-section data we expect $\alpha \rightarrow 0$ as well as
$\gamma\rightarrow -6$, and therefore, $s^{6+\gamma}\rightarrow 1$ and $s^{\alpha}\rightarrow 1$ indicating the saturation of the
Froissart-Martin bound. However, at the present day energy scale, we do not expect $\alpha=0$ and so even really small values of
$\alpha$ will give a slow increase with $s$ for the total cross-section when compared with the "full" Froissart-Martin
bound. One consider here the $\alpha$-parameter as some kind of ``correction term'' in the Froissart-Martin bound at low energies and it may represents some physical mechanisms that avoid the total cross-section saturation.

As mentioned above, $\alpha$ may be treated as a correction in the Froissart-Martin bound at low energy. However, at (finite) high energies all the second term in the right hand side of (\ref{pp}) and (\ref{pbp}) will behave as some power of $s$
\begin{eqnarray}
\label{pomeron} \frac{\pi\beta}{\sqrt{2}s^{\alpha}m_{\pi}^2}
\ln^2s \rightarrow cs^{p}.
\end{eqnarray}

\noindent where $c$ is a convenient constant and $p$ the effective index.

In order to apply the results obtained above to analyze the total cross-section data, we notice the important result assuring that
if the Froissart-Martin is reached then the difference between $pp$ and $\bar{p}p$ total cross-section goes as
\cite{matthiae,eden,varios2}
\begin{eqnarray}
\nonumber \Delta \sigma_{\footnotesize\mbox{tot}} =
\sigma_{\footnotesize\mbox{tot}}^{pp}-\sigma_{\footnotesize\mbox{tot}}^{\bar{p}p}\leq
c\frac{\sigma_{\footnotesize\mbox{tot}}^{pp}
+\sigma_{\footnotesize\mbox{tot}}^{\bar{p}p}}{\ln s},
\end{eqnarray}

\noindent which means that the difference can increases at most as $\ln s$ and even in this case,
\begin{eqnarray}
\nonumber
\frac{\sigma_{\footnotesize\mbox{tot}}^{pp}}{\sigma_{\footnotesize\mbox{tot}}^{\bar{p}p}}\rightarrow 1,
\end{eqnarray}

\noindent as $s \rightarrow \infty$. This is the revised version of the Pomeranchuk theorem and in order to obey this formal
result, we must impose the following constraints
\begin{eqnarray}
\label{constraints} \delta=\bar{\delta},\hspace{0.3cm} \gamma=\bar{\gamma},\hspace{0.3cm}
\alpha=\bar{\alpha},\hspace{0.3cm} \beta=\bar{\beta}.
\end{eqnarray}

Nevertheless, the energies achieved in the collision processes allow to release these constraints and therefore, as a first step,
we adopt in the fitting procedure four independent parameters to $pp$ and four to $\bar{p}p$.

\section{\label{sec4}Real Part of the Elastic Scattering Amplitude and $\rho$ Parameter}
In this section, we shall obtain the real part of the forward elastic scattering amplitude. In particular, we obtain $\rho(s)$, the ratio of the real to imaginary part of the forward scattering amplitude in the high energy domain.

The optical theorem connects the total cross-section to imaginary part of the elastic scattering at $q^2=0$. On the other hand,
since the imaginary part of the elastic scattering amplitude may be connected to the real part using dispersion relation one
concludes that forward elastic scattering amplitude possess the same free parameters, except for the free-parameters coming from the
subtraction terms of the dispersion relation, of the total cross-section. Connections
between real and imaginary parts of the forward scattering amplitude have been widely investigated by dispersion relation in
both integral and derivative forms (we use here the derivative form). Several papers have been devoted to calculations of the
real parts of the $pp$ and $\bar{p}p$ forward scattering amplitude, using a wide variety of dispersion relations and
different representations for the energy dependence of the imaginary part of the scattering amplitude at $q^2 = 0$. Here, to
obtain the real part of the elastic scattering amplitude, we use the approach adopted in \cite{sdc,jackson}. We define two
auxiliary functions (crossing between $pp$ and $\bar{p}p$)
\begin{eqnarray}
\nonumber \frac{\mathrm{Im}f_{\pm}}{s}=\frac{1}{2}\left[\frac{\mathrm{Im}F_{pp}}{s}\pm
\frac{\mathrm{Im}F_{\bar{pp}}}{s}\right],
\end{eqnarray}

\noindent and relate these functions with the elastic scattering amplitude $F_{pp/\bar{p}p}=f_+ \pm f_-$,
where $f_+=\mathrm{Re}f_++i\mathrm{Im}f_+$ and $f_-=\mathrm{Re}f_-+i\mathrm{Im}f_-$ and $f_{\pm}$ are convenient
amplitudes. In the forward direction, the derivative dispersion relation for even (+) and odd (-) amplitudes are expressed in
terms of a tangent operator and in the case of one subtraction (equal to two subtractions in the even case) they are given by \cite{avilamenon}
\begin{eqnarray}
\nonumber \frac{\mathrm{Re}f_+(s,0)}{s}=\frac{k}{s}+\tan\left[\frac{\pi}{2}\frac{d}{d\ln s}\right]
\frac{\mathrm{Im}f_+(s,0)}{s},
\end{eqnarray}

\begin{eqnarray}
\nonumber \frac{\mathrm{Re}f_-(s,0)}{s}=\tan\left[\frac{\pi}{2}\left(1+\frac{d}{d\ln s}\right)\right]
\frac{\mathrm{Im}f_-(s,0)}{s}.
\end{eqnarray}

On the other hand, Bronzan, Kane and Sukhatme \cite{bks} obtained derivative dispersion relations (without subtraction constant) using an integration parameter $\nu$. This parameter was used as free fit parameter or considered as constant over years \cite{avilamenon}.
\begin{eqnarray}
\nonumber \frac{\mathrm{Re}f_+(s,0)}{s^{\nu}}=\tan\left[\frac{\pi}{2}\left(\nu-1+\frac{d}{d\ln s}\right)\right]
\frac{\mathrm{Im}f_+(s,0)}{s^{\nu}},
\end{eqnarray}

\begin{eqnarray}
\nonumber \frac{\mathrm{Re}f_-(s,0)}{s^{\nu}}=\tan\left[\frac{\pi}{2}\left(\nu+\frac{d}{d\ln s}\right)\right]\frac{\mathrm{Im}f_-(s,0)}{s^{\nu}},
\end{eqnarray}

\noindent and Kang and Nicolescu \cite{kangnicolescu} obtained derivative dispersion relations without $\nu$ parameter (without subtraction constant) and
\begin{eqnarray}
\nonumber \frac{\mathrm{Re}f_+(s,0)}{s}=\left[\frac{\pi}{2}\frac{d}{d\ln s}+\frac{1}{3}\left(\frac{\pi}{2}\frac{d}{d\ln s} \right)^3+ \frac{2}{5}\left(\frac{\pi}{2}\frac{d}{d\ln s}\right)^5+...\right]
\frac{\mathrm{Im}f_+(s,0)}{s},
\end{eqnarray}

\begin{eqnarray}
\nonumber \frac{\mathrm{Re}f_-(s,0)}{s}=-\frac{2}{\pi}\int{\left[1-\left(\frac{\pi}{2}\frac{d}{d\ln s}\right)^2-\frac{1}{45}\left(\frac{\pi}{2}\frac{d}{d\ln s}\right)^4-...\right]}
\frac{\mathrm{Im}f_-(s,0)}{s}d\ln s.
\end{eqnarray}

\'Avila and Menon \cite{avilamenon} obtained the same result as
Bronzan, Kane and Sukhatme without $\nu$ parameter, which is
equivalent to consider $\nu=1$ in Bronzan, Kane and Sukhatme
results. The odd case of Kang and Nicolescu can be obtained from
\'Avila and Menon expressions replacing
$\tan\left[\frac{\textstyle \pi}{\textstyle
2\rule{0pt}{8.5pt}}\left(1+\frac{\textstyle d}{\textstyle d\ln
s\rule{0pt}{8.5pt}}\right)\right] \rightarrow
-\cot\left[\frac{\textstyle \pi}{\textstyle
2\rule{0pt}{8.5pt}}\frac{\textstyle d}{\textstyle d\ln
s\rule{0pt}{8.5pt}}\right]$ and expanding the series around the
origin in \'Avila and Menon derivative dispersion relations
\cite{avilamenon}. Furthermore, results of \'Avila and Menon are
equivalent to Bronzan, Kane and Sukhatme considering the
particular case $\nu=1$.

It is important to stress that replacement
$\tan\left[\frac{\textstyle \pi}{\textstyle
2\rule{0pt}{8.5pt}}\left(1+\frac{\textstyle d}{\textstyle d\ln
s\rule{0pt}{8.5pt}}\right)\right] \rightarrow
-\cot\left[\frac{\textstyle \pi}{\textstyle
2\rule{0pt}{8.5pt}}\frac{\textstyle d}{\textstyle d\ln
s\rule{0pt}{8.5pt}}\right]$ may present some theoretical flow. As
known, the first term of the $\cot(x)$ expansion presents the
application of the inverse of the derivative operator.
Mathematically speaking, let $\Pi$ a linear operator. Its inverse,
$\Pi^{-1}$, is by definition ($H(x)$ is a non-zero analytic
function)
\begin{eqnarray}
\nonumber\Pi[\Pi^{-1}H(x)]=\Pi^{-1}[\Pi{H(x)}]=H(x).
\end{eqnarray}

Furthermore, $\Pi^{-1}$ is also linear \cite{goldberg}. Of course,
we consider $x$ as an element of a linear space with a
well-defined norm. In the context of the operatorial algebra,
there exist $\Pi^{-1}$ if and only if to $m>0$
\begin{eqnarray}
\label{def} m\|H(x)\|\leq\|\Pi{H(x)}\|.
\end{eqnarray}

To the especific case where $\Pi\equiv\frac{\textstyle
d}{\textstyle d\ln s\rule{0pt}{8.5pt}}$ is always possible obtain
$m$ to satisfy the above result. However, we must take in to
account the possible divergence result, i.e.,
\begin{eqnarray} \nonumber \frac{d}{d\ln s}H(s)=0,
\end{eqnarray}

\noindent and therefore $1/(d/d\ln s)\rightarrow \infty$. The
above result may be viewed as a possible limitation to the
functional form of the imaginary part of the forward amplitude: it
cannot be constant in $d/d\ln s$. Of course, this result is valid
only if the formal procedure presented above is correct.

It has also been demonstrated by Fischer and Kol\'a\v{r} \cite{kolar1,kolar2,kolar3} that at high energies the above even tangent
operator can be replaced by its first order expansion. The odd representation also can be replaced by its first term \cite{kolar3} by using the notation of Kang and Nicolescu. However, \'Avila and Menon and Kang and Nicolescu representations are equivalent by the transformation showed above.
Therefore, we retain here, for the sake of simplicity, only the first term of tangent series expansion in the \'Avila and Menon representation for both even and odd derivative dispersion relations.
\begin{eqnarray}
\nonumber \frac{\mathrm{Re}f_+(s,0)}{s}=\frac{k}{s}+\frac{\pi}{2}\frac{d}{d\ln s}\frac{\mathrm{Im}f_+(s,0)}{s},
\end{eqnarray}

\begin{eqnarray}
\nonumber \frac{\mathrm{Re}f_-(s,0)}{s}=\frac{\pi}{2}\left(1+\frac{d}{d\ln s}\right)\frac{\mathrm{Im}f_-(s,0)}{s}.
\end{eqnarray}

This approach was recently used presenting good statistical results \cite{sdc}. Of course, other terms of the tangent series expansion could be considered as perturbations of the first (leading) term.

The real part of the scattering amplitude is given by
\begin{eqnarray}
\nonumber \frac{\mathrm{Re}F_{pp/{\bar{p}p}}(s,0)}{s}=\frac{\mathrm{Re}f_+(s,0)}{s}\pm \frac{\mathrm{Re}f_-(s,0)}{s}.
\end{eqnarray}

The most general versions of local quantum field theory lead to proving dispersion relation and, more generally, analyticity
properties in two variables in a rather large domain, if one makes use of the positivity properties of the absorptive part of the
scattering amplitude. Empirically, over many years, dispersion relation had always been consistent with measured data for
energies reached by fixed target machines (e.g. pion nucleon scattering) or colliders (ISR and $\mathrm{Sp\bar{p}S}$
colliders). Therefore, considering the above formalism we obtain ($pp$ and $\bar{p}p$)


\begin{eqnarray}
\label{jac7}
\frac{\mathrm{Re}F_{pp}(s,0)}{s}=\frac{k}{s}+\frac{\pi}{2}\left[-\frac{\delta(6+\gamma)}{s^{6+\gamma}}+\frac{\pi\beta\sqrt{2}}{m_{\pi}^2s^{\alpha}}\left(\ln s-\frac{\alpha\ln^2 s}{2}\right)\right]+\frac{\pi}{4}[\sigma_{\footnotesize\mbox{tot}}^{pp}-\sigma_{\footnotesize\mbox{tot}}^{\bar{p}p}],
\end{eqnarray}

\begin{eqnarray}
\label{jac8}
\frac{\mathrm{Re}F_{\bar{p}p}(s,0)}{s}=\frac{k}{s}+\frac{\pi}{2}\left[-\frac{\bar{\delta}(6+\bar{\gamma})}{s^{6+\bar{\gamma}}}+\frac{\pi\bar{\beta}\sqrt{2}}{m_{\pi}^2s^{\bar{\alpha}}}\left(\ln s-\frac{\bar{\alpha}\ln^2 s}{2}\right)\right]-\frac{\pi}{4}[\sigma_{\footnotesize\mbox{tot}}^{pp}-\sigma_{\footnotesize\mbox{tot}}^{\bar{p}p}],
\end{eqnarray}

\noindent where the crossing property allows the simultaneous analysis to $pp$ and $\bar{p}p$ scattering amplitude. Notice that
$pp$ and $\bar{p}p$ forward scattering amplitudes interact with each to other by means the crossing property.

At high energies is expected that the imaginary part of the elastic scattering amplitude dominate the real part, i.e., the
absorptive part will represents the whole scattering amplitude. Therefore, a simple measure of how fast the imaginary part
increases is given by $\rho$, the ratio of the real to imaginary part of the forward scattering amplitude in the high energy domain
\begin{eqnarray}
\nonumber \rho(s)=\frac{\mathrm{Re}F(s,0)}{\mathrm{Im}F(s,0)},
\end{eqnarray}

\noindent and using (\ref{pp}), (\ref{pbp}), (\ref{jac7}) and
(\ref{jac8}) we can explicitly write $\rho(s)$ as ($pp$ and
$\bar{p}p$)
\begin{eqnarray}
\label{rhopp} \rho^{pp}(s)=\frac{\frac{\textstyle k}{\textstyle
s\rule{0pt}{8.5pt}}+\frac{\textstyle \pi}{\textstyle
2\rule{0pt}{8.5pt}}\left[-\frac{\textstyle
\delta(6+\gamma)}{\textstyle s^{6+\gamma}\rule{0pt}{8.75pt}}+
\frac{\textstyle \pi\beta\sqrt{2}}{\textstyle
m_{\pi}^2s^{\alpha}\rule{0pt}{8.75pt}}\left(\ln s-\frac{\textstyle
\alpha}{\textstyle 2\rule{0pt}{8.5pt}}\ln^2
s\right)\right]+\frac{\textstyle \pi}{\textstyle
4\rule{0pt}{8.5pt}}[\sigma_{\footnotesize\mbox{tot}}^{pp}-
\sigma_{\footnotesize\mbox{tot}}^{\bar{p}p}]} {\frac{\textstyle
\delta\rule{0pt}{8.75pt}}{\textstyle
s^{6+\gamma}\rule{0pt}{8.75pt}} + \frac{\textstyle
\pi\beta\rule{0pt}{8.75pt}}{\textstyle
\sqrt{2}m_{\pi}^2s^{\alpha}\rule{0pt}{8.75pt}} \ln^2s},
\end{eqnarray}
\vspace*{-0.3cm}
\begin{eqnarray}
\label{rhopbp} \rho^{\bar{p}p}(s)=\frac{\frac{\textstyle
k}{\textstyle s\rule{0pt}{8.5pt}}+\frac{\textstyle \pi}{\textstyle
2\rule{0pt}{8.5pt}}\left[-\frac{\textstyle
\bar{\delta}(6+\bar{\gamma})} {\textstyle
s^{6+\bar{\gamma}}\rule{0pt}{9.25pt}}+\frac{\textstyle
\pi\bar{\beta}\sqrt{2}}{\textstyle
m_{\pi}^2s^{\bar{\alpha}}\rule{0pt}{9.25pt}}\left(\ln
s-\frac{\textstyle \bar{\alpha}}{\textstyle
2\rule{0pt}{8.5pt}}\ln^2 s\right)\right]-\frac{\textstyle
\pi}{\textstyle
4\rule{0pt}{8.5pt}}[\sigma_{\footnotesize\mbox{tot}}^{pp}-
\sigma_{\footnotesize\mbox{tot}}^{\bar{p}p}]} { \frac{\textstyle
\bar{\delta}\rule{0pt}{8.75pt}}{\textstyle
s^{6+\bar{\gamma}}\rule{0pt}{9.25pt}} + \frac{\textstyle
\pi\bar{\beta}\rule{0pt}{8.75pt}}{\textstyle
\sqrt{2}m_{\pi}^2s^{\bar{\alpha}}\rule{0pt}{9.25pt}} \ln^2s}.
\end{eqnarray}



The only free parameter in the above results is the subtraction constant $k$. It is not difficult to see that
$\rho(s)=\rho\rightarrow \ln^{-1}s$ as $s\rightarrow \infty$. In the next sections, we shall show both individual and
simultaneous fit results to $\sigma_{\footnotesize\mbox{tot}}$ and $\rho$ parameter.

\section{\label{sec3}Individual Fit Results for $\sigma_{\footnotesize\mbox{tot}}$ and $\rho$}
Available experimental data on $\sigma_{\footnotesize\mbox{tot}}$
for proton-proton and antiproton-proton scattering were fitted by
functions (\ref{pp}) and (\ref{pbp}) respectively. The database
compiled by Particle Data Group has become a standard references
and the corresponding computer readable files are used \cite{pdg}.
The statistic and systematic errors are added linearly in our
analysis just as well as for some early analysis (see, for example
\cite{menon,okorokov1,okorokov2}).

\begin{figure}[t!]
\centering{\includegraphics[width=17.0cm,height=17.0cm]{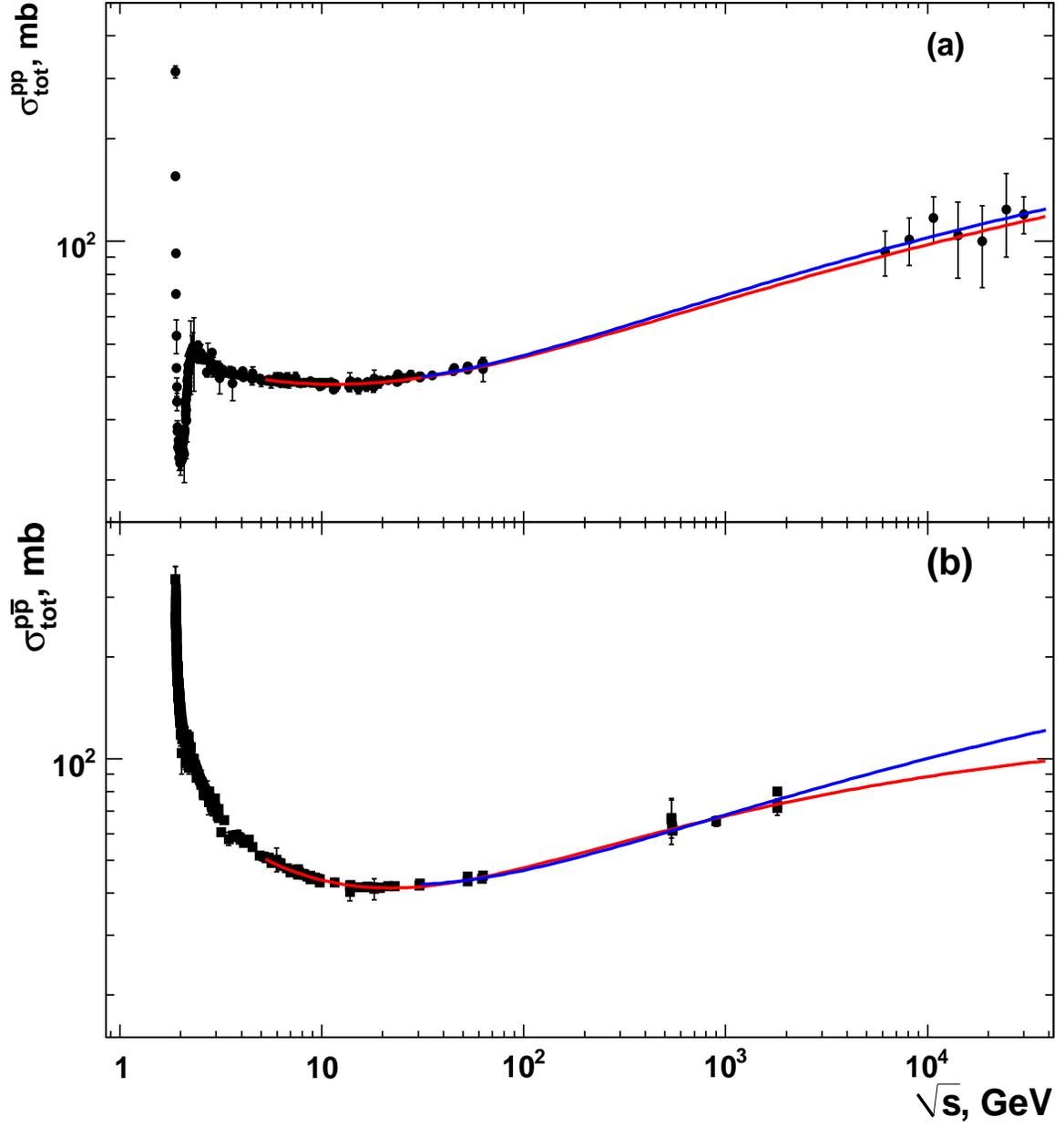}}
\caption{(color online) Total cross-section energy dependence for
proton-proton (a) and antiproton-proton (b) collisions. The red
lines correspond to individual fits at
$\sqrt{s_{\footnotesize\mbox{min}}}=5$, the blue lines - at
$\sqrt{s_{\footnotesize\mbox{min}}}=30$. Experimental data are
from \cite{pdg}.} \label{fig:fig-sigma1}
\end{figure}

As seen the suggested parameterizations are valid only for $s\geq
s_{\footnotesize\mbox{min}}$, where
$s_{\mbox{\footnotesize{min}}}$ is some empirical low boundary (in
GeV). During the analysis procedure we decrease the
$s_{\mbox{\footnotesize{min}}}$ value as much as possible in order
to describe the wider energy domain with statistically reasonable
fit quality. This logic is used in all our fit procedure below.
Thus the fits have been made only at $s\geq
s_{\footnotesize\mbox{min}}$ for different values of low boundary.
The fitting parameter values are presented in Table
\ref{fig:tabela1} for $pp$ and in Table \ref{fig:tabela2} for
$\bar{p}p$. Figure \ref{fig:fig-sigma1} shows the experimental
energy dependence for fit for $pp$ (Fig. \ref{fig:fig-sigma1}a),
$\bar{p}p$ (Fig. \ref{fig:fig-sigma1}b) and corresponding fits at
$\sqrt{s_{\footnotesize\mbox{min}}}=5$ (red lines),
$\sqrt{s_{\footnotesize\mbox{min}}}=30$ (blue lines).

Parametrization (\ref{pp}) agree to experimental data very well
for any $\sqrt{s_{\footnotesize\mbox{min}}}$ under study. One
needs to emphasize that (\ref{pp}) allows us to describe
experimental points at $\sqrt{s_{\footnotesize\mbox{min}}}=3$ very
well. We do not see clear energy dependence of $\alpha$ and
$\beta$ within error bars but these parameters are very small and
agree to expected values (Table \ref{fig:tabela1}). One can see
the statistically acceptable fit qualities are obtained for
$\sqrt{s_{\footnotesize\mbox{min}}}=5-15$ only for $\bar{p}p$
(Table \ref{fig:tabela2}). Fit function (\ref{pbp}) agree to
experimental points at qualitative level well for other low
boundary values under study. But fit quality is statistically
unacceptable at lowest $\sqrt{s_{\footnotesize\mbox{min}}}$ and
some worse for $\sqrt{s_{\footnotesize\mbox{min}}}=20-30$
\footnote{Values of fit parameters are equals at
$\sqrt{s_{\footnotesize\mbox{min}}}=25$ and for
$\sqrt{s_{\footnotesize\mbox{min}}}=30$ because of absence of
experimental $\bar{p}p$ points in the range of collision energies
$\sqrt{s}= 25-30$ GeV.}. The $\bar{\alpha}$ as well as
$\bar{\beta}$ some decrease up to
$\sqrt{s_{\footnotesize\mbox{min}}}=10$ and remain almost
constants within errors at larger
$\sqrt{s_{\footnotesize\mbox{min}}}$. On the other hand it seems
small values of $\alpha$, $\bar{\alpha}$ agree with predicted
behavior $\sigma_{\footnotesize\mbox{tot}}^{NN}\propto \ln^2 s$ at
high energies. One needs to emphasize the constraints
(\ref{constraints}) are valid for fits at
$\sqrt{s_{\footnotesize\mbox{min}}}\geq 15$ within errors.

\begin{figure}[t!]
\centering{\includegraphics[width=17.0cm,height=17.0cm]{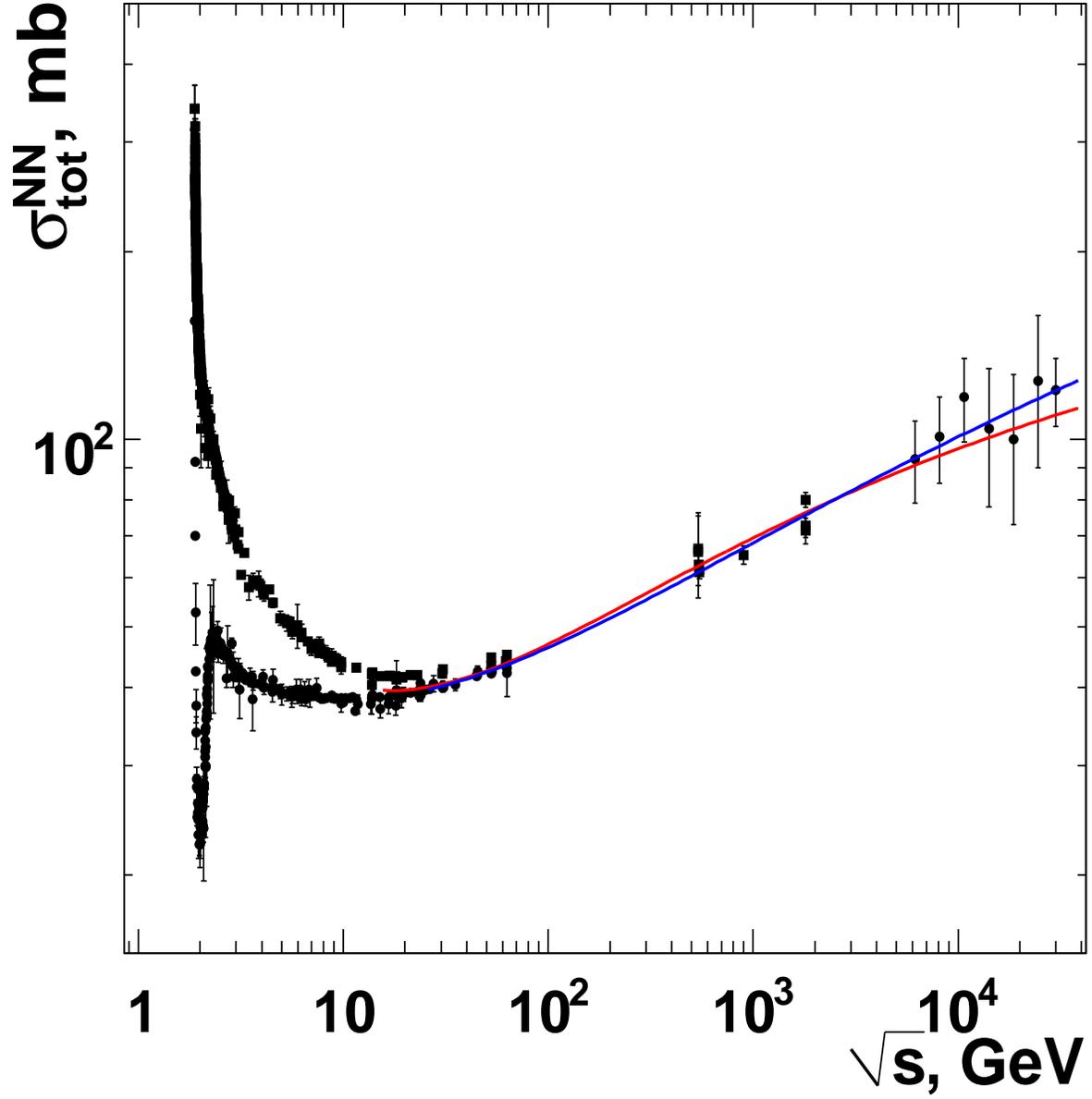}}
\caption{(color online) Total cross-section energy dependence for
nucleon-nucleon collisions. The red line corresponds to the fit at
$\sqrt{s_{\footnotesize\mbox{min}}}=15$, the blue line - at
$\sqrt{s_{\footnotesize\mbox{min}}}=25$. Experimental points are
indicated as {\large$\bullet$} ($\blacksquare $) for $pp$
($\bar{p}p$).Experimental data are from \cite{pdg}.}
\label{fig:fig-sigma2}
\end{figure}

In accordance with (\ref{pp}) and (\ref{pbp}) total cross-section
energy dependence is described by the same functional form for
$pp$ and $\bar{p}p$ reactions. Thus one can make the quantitative
analysis of common data ensemble for
$\sigma_{\footnotesize\mbox{tot}}^{pp}$ and
$\sigma_{\footnotesize\mbox{tot}}^{\bar{p}p}$ taking into account
above results. The fitting parameter values are presented in Table
\ref{fig:tabela3} for $NN$ scattering. As seen the fit quality is
statistically unacceptable for
$\sqrt{s_{\footnotesize\mbox{min}}}=15$ despite of equality of fit
parameters for $pp$ and $\bar{p}p$. Our quantitative analysis
demonstrates that experimental data for $pp$ and $\bar{p}p$ total
cross-section can be described by one function at
$\sqrt{s_{\footnotesize\mbox{min}}} \geq 25$ with statistically
reasonable quality. Fig. \ref{fig:fig-sigma2} shows the
experimental energy dependence for
$\sigma_{\footnotesize\mbox{tot}}^{NN}$ and corresponding fits for
two various $\sqrt{s_{\footnotesize\mbox{min}}}$: fit at low
energy limit for simultaneous study of proton-proton and
antiproton-proton scattering
$\sqrt{s_{\footnotesize\mbox{min}}}=15$ is shown by red line, blue
line corresponds to the fit with best quality at
$\sqrt{s_{\footnotesize\mbox{min}}}=25$.

If we fix the set parameters we are capable to obtain theoretical
predictions for $\sigma_{\footnotesize\mbox{tot}}$ at the RHIC,
LHC and at the future HC energies. There are several predictions
for the total cross-section covering a wide range, from 80 mb up
to 230 mb and at Tevatron ($\sqrt{s}=1.8$ TeV)
$\sigma_{\footnotesize\mbox{tot}}$ is ill-determined,
$\sigma_{\mbox{\footnotesize{tot}}}=(71.42\pm2.41)$ mb for E710
Collaboration and
$\sigma_{\mbox{\footnotesize{tot}}}=(80.03\pm2.24)$ mb according
to CDF Collaboration. Using $\sqrt{s_{\footnotesize\mbox{min}}}=5$
as the minimum low boundary value with statistically acceptable
fit quality for $\bar{p}p$ data one obtains
$\sigma_{\mbox{\footnotesize{tot}}}^{\bar{p}p}=(73.22 \pm 8.30)$
mb as a prediction for $\sqrt{s}=1.8$ TeV.

On the other hand, for $\sqrt{s}=546$ GeV (closest to the maximum
RHIC energy), CDF Collaboration obtains
$\sigma_{\mbox{\footnotesize{tot}}}=(61.3\pm1.0)$ mb, UA4
Collaboration $\sigma_{\mbox{\footnotesize{tot}}}=(61.9\pm1.5)$
mb. Using $\sqrt{s_{\footnotesize\mbox{min}}}=5$ one obtains
$\sigma_{\mbox{\footnotesize{tot}}}^{\bar{p}p}=(61.93\pm 6.17)$ mb
for $\sqrt{s}=546$ GeV.

At $\sqrt{s}=14$ TeV, Block \textit{et al} predict
$\sigma_{\mbox{\footnotesize{tot}}}^{pp}=(108.0\pm 3.4)$ mb
\cite{bghp1} and Bourrely, Soffer and Wu predict
$\sigma_{\mbox{\footnotesize{tot}}}^{pp}=103.5$ mb \cite{bsw}. In
Table \ref{fig:tabela4} our predictions are shown for the $pp$
total cross-section at LHC energy ranges, the future HC and at
RHIC. To summarize, our predictions indicate
$\sigma_{\mbox{\footnotesize{tot}}}\simeq 100.44$ mb at
$\sqrt{s}=14$ TeV, $\sigma_{\mbox{\footnotesize{tot}}}\simeq
95.99$ mb at $\sqrt{s}=10$ TeV, and
$\sigma_{\mbox{\footnotesize{tot}}}\simeq 91.35$ mb at
$\sqrt{s}=7$ TeV (possible LHC new energies).

The extraction of the $pp$ total cross-section from cosmic-ray
experiments is based on the determination of proton-air production
cross-section. The procedure is model-dependent
\cite{cosmic-ray1}. The Fly's Eye Collaboration \cite{cosmic-ray2}
using the Geometrical Scaling model obtain to $pp$ at
$\sqrt{s}=30$ TeV $\sigma_{\mbox{\footnotesize{tot}}}=(122\pm11)$
mb. At same energy one obtains
$\sigma_{\mbox{\footnotesize{tot}}}=(110.69 \pm 19.32)$ mb at
$\sqrt{s_{\footnotesize\mbox{min}}}=3$ as well as lowest energy
boundary with statistically acceptable fit quality for $pp$
scattering. On the other hand, Gaisser, Sukhatme and Yodh
\cite{gaisser} using the Fly's Eye result and the Chou-Yang
prescription for the slope \cite{chouyang} obtained at
$\sqrt{s}=40$ TeV for $pp$ interaction
$\sigma_{\mbox{\footnotesize{tot}}}=175_{-27}^{+40}$ mb. At same
energy one obtains
$\sigma_{\mbox{\footnotesize{tot}}}=(114.60\pm20.55)$ mb. Of
course, results from cosmic-rays presents a theoretical bias as
noted by Nikolaev \cite{nikolaev}. When the Akeno's result at
$\sqrt{s}=30$ TeV ($\sim 92$ mb) was corrected the corresponding
$\sigma_{\mbox{\footnotesize{tot}}}$ increase (amazing) on $\sim
30$ mb.

Even at higher energies than that accessible in the present
collider experiments, Baltrusaitis \textit{et al} predicts
$\sigma_{\mbox{\footnotesize{tot}}}^{pp}=(120\pm 15)$ mb at
$\sqrt{s}=30$ TeV \cite{cosmic-ray2}. At same energy our
prediction using the parameters obtained starting at
$\sqrt{s_{\footnotesize\mbox{min}}}=3$ is
$\sigma_{\mbox{\footnotesize{tot}}}^{pp}=(110.69\pm 19.32)$ mb.
Finally, Block and Cahn \cite{blockcahn} obtain at very high
energies ($\sqrt{s}=500$ TeV)
$\sigma_{\mbox{\footnotesize{tot}}}^{pp}=(316.00\pm 5.46)$ mb and
our prediction to the same energy is
$\sigma_{\mbox{\footnotesize{tot}}}^{pp}=(149.25\pm 32.85)$ mb,
almost half. It is important to stress that the fit qualities for
(\ref{pp}) and (\ref{pbp}) are better for standard PDG
parametrization at $\sqrt{s}\geq$ 5 GeV and suggested
parameterizations allows to represent nucleon-nucleon
$\sigma_{\mbox{\footnotesize{tot}}}$ energy dependencies in wide
energy domain.

The experimental data for ratio of the real to imaginary part of
the forward scattering amplitude are some poorer than that for
total cross-sections especially in high energy domain. Below in
this paragraph individual fits under study only for $\rho$
parameter energy dependence in $pp$ and $\bar{p}p$ scattering.

The fitting functions (\ref{rhopp}) and (\ref{rhopbp}) are used
for $\rho^{pp}(s)$ and for $\rho^{\bar{p}p}(s)$ respectively.

There are two possible approaches for individual fit of $\rho$
parameter for each interaction under considered, namely, i) the
only one free parameter $k$ is used in fitting functions
(\ref{rhopp}) and (\ref{rhopbp}) for $\rho^{pp}(s)$ and for
$\rho^{\bar{p}p}(s)$ but values of all remain parameters are fixed
by the individual fits of total cross-sections at corresponding
$\sqrt{s_{\footnotesize\mbox{min}}}$; and ii) all parameters in
(\ref{rhopp}) and (\ref{rhopbp}) are free, results of individual
fits for $\sigma_{\mbox{\footnotesize{tot}}}^{pp}$ and
$\sigma_{\mbox{\footnotesize{tot}}}^{\bar{p}p}$ are not taken into
account at all. The fit qualities are reasonable for any
$\sqrt{s_{\footnotesize\mbox{min}}}$ both for $pp$ (Table
\ref{fig:tabela5}, part i) and $\bar{p}p$ (Table
\ref{fig:tabela6}, part i) if the corresponding parameter values
are fixed based on the above results for total cross-sections and
only $k$ parameter is free. Moreover, $\chi^{2}/\mbox{n.d.f.}$ are
statistically acceptable at $\sqrt{s_{\footnotesize\mbox{min}}}
> 5$ for $pp$ and at some values of
$\sqrt{s_{\footnotesize\mbox{min}}}$ for $\bar{p}p$ in the
framework of i) fit approach. On the other hand the precision of
$k$ parameter is some poor especially for $\bar{p}p$ interaction.
The fit with all free parameters was studied also. The results for
fit approach ii) are shown in Table \ref{fig:tabela5} (part ii)
for $pp$ scattering and in Table \ref{fig:tabela6} (part ii) for
$\bar{p}p$ reaction\footnote{Values of fit parameters are equals
at $\sqrt{s_{\footnotesize\mbox{min}}}=5$ and for
$\sqrt{s_{\footnotesize\mbox{min}}}=10$ because of absence of
experimental $\bar{p}p$ points in the range of collision energies
$\sqrt{s}= 5-10$ GeV.}. For approach ii) high energy boundaries
for individual fits of $\rho$ have to be decreased because of
large number of free parameters and small data samples for high
energy domain. In this case the fit qualities are statistically
acceptable for all $s_{\footnotesize\mbox{min}}$ under study both
for $pp$ and $\bar{p}p$ interactions.

\begin{figure}[t!]
\centering{\includegraphics[width=17.0cm,height=17.0cm]{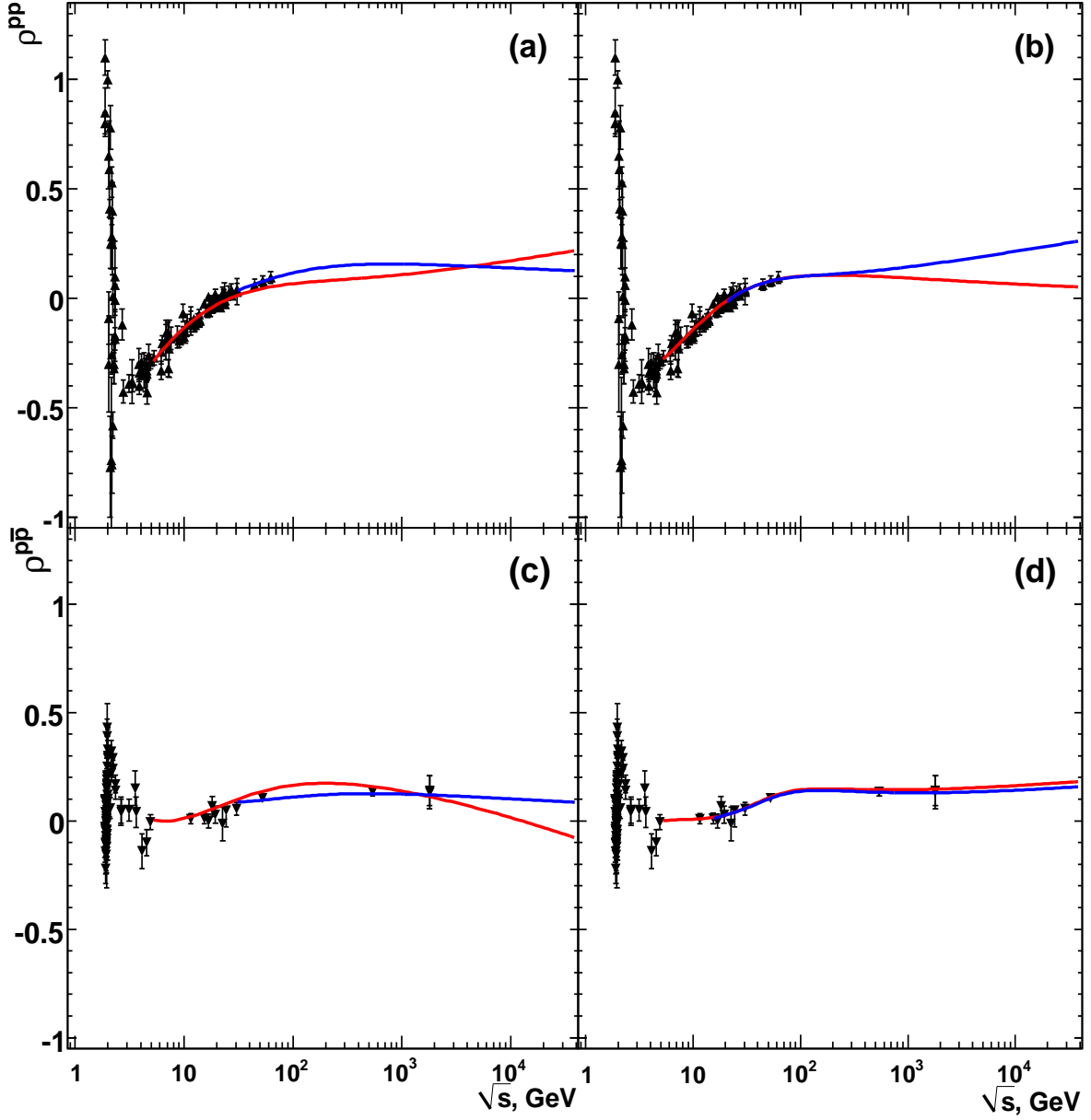}}
\caption{(color online) The $\rho$ parameter energy dependence for
proton-proton (a,b) and antiproton-proton (c,d) collisions.
Results obtained in the framework of i) approach for individual
fit are shown on (a,c) and for approach ii) - on (b,d). The red
lines correspond to the individual fits at
$\sqrt{s_{\footnotesize\mbox{min}}}=5$ for any approaches. Blue
lines correspond to the individual fits at
$\sqrt{s_{\footnotesize\mbox{min}}}=30$ for i) approach (a,c) and
for approximation curves obtained in the framework of ii) approach
at $\sqrt{s_{\footnotesize\mbox{min}}}=20 ~(15)$ for
$pp~(\bar{p}p)$ respectively (b,d). Experimental data are from
\cite{pdg}.} \label{fig:fig-rho1}
\end{figure}

As seen from Table \ref{fig:tabela5} and from Table
\ref{fig:tabela6} precision of fit parameters is some poorer than
that for total cross-section fits for both approaches i) and ii).
Fig. \ref{fig:fig-rho1} shows the experimental energy dependence
of $\rho$ parameter for $pp$ (Fig. \ref{fig:fig-rho1}a,b), for
$\bar{p}p$ (Fig. \ref{fig:fig-rho1}c,d) elastic scattering and
corresponding fits at various values of low boundary. Fits at
$\sqrt{s_{\footnotesize\mbox{min}}}=5$ are shown by red lines at
all figures, fits at low boundary
$\sqrt{s_{\footnotesize\mbox{min}}}=30$ are available for i)
approach only and shown by blue lines at Fig.
\ref{fig:fig-rho1}a,c. The same color lines correspond to the
$\sqrt{s_{\footnotesize\mbox{min}}}=20$ (15) for $pp$ ($\bar{p}p$)
reaction for ii) approach at Fig. \ref{fig:fig-rho1}b,d. In the
framework of i) approach fit curves show a noticeable increasing
/decreasing of $\rho$ parameter value for $pp$ / $\bar{p}p$
interactions respectively for
$\sqrt{s_{\footnotesize\mbox{min}}}=3-10$ at high energies. The
changing of absolute value of $\rho$ is weaker at increasing of
$\sqrt{s_{\footnotesize\mbox{min}}}$ (Fig. \ref{fig:fig-rho1}a,c).
At higher values of low boundary fit curves for $pp$ demonstrate
some maximum value in the $\sqrt{s}\sim 100$ GeV region but curves
for $\bar{p}p$ elastic scattering show the much more flat behavior
than that for $pp$ reaction. Value of the maximum for $pp$
reaction increases with increasing of low boundary for fitted
energy domain at $\sqrt{s_{\footnotesize\mbox{min}}}=15-25$. As
seen from Fig. \ref{fig:fig-rho1}b the situation with increasing
of low boundary is opposite in the framework of ii) approach for
$pp$. The fit curves at Fig. \ref{fig:fig-rho1}d show increasing
of $\rho$ parameter for $\bar{p}p$ interaction at any
$\sqrt{s_{\footnotesize\mbox{min}}}$ under considered for ii)
approach. It seems the model curves obtained in the framework of
i) approach with one $k$ free parameter show more reasonable and
expected behaviour in the physics sense at
$\sqrt{s_{\footnotesize\mbox{min}}} \geq 15$ at least than that in
the framework of ii) approach. One can see a very small values for
$\beta$ at two highest values of
$\sqrt{s_{\footnotesize\mbox{min}}}$ and for $\bar{\beta}$
parameters in the case of individual fits of $\rho^{pp/\bar{p}p}$
energy dependence. In the framework of ii) approach constraints
(\ref{constraints}) are valid for all fit parameters with
exception of $(\delta,~\bar{\delta})$ pair at highest
$\sqrt{s_{\footnotesize\mbox{min}}}=15$ at which comparison is
available for parameter values for $pp$ and $\bar{p}p$ reactions.
But one needs to emphasize that qualities for individual fits are
well enough for $\sqrt{s_{\footnotesize\mbox{min}}}=3$ even and
suggested formulas (\ref{rhopp}), (\ref{rhopbp}) allow us to get a
good agreement between model curves and experimental data in wide
energy domain $\sqrt{s} \geq 3$ GeV.

\section{\label{secSimult}Simultaneous Fit Results for $\sigma_{\mbox{\footnotesize{tot}}}$ and $\rho$}
\begin{figure*}[t!]
\centering{\includegraphics[width=17.0cm,height=17.0cm]{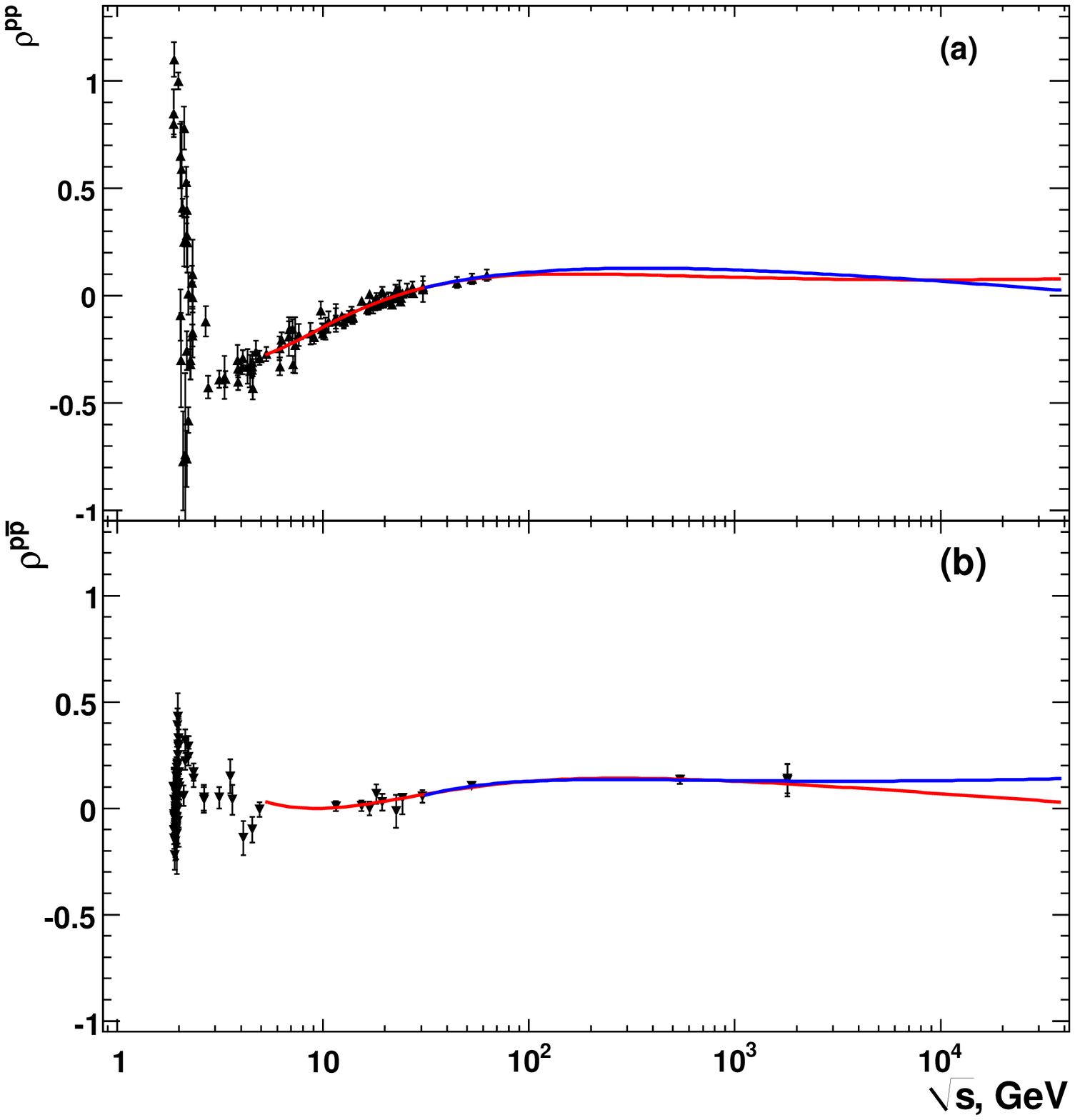}}
\caption{(color online) The $\rho$ parameter energy dependence for
proton-proton (a) and antiproton-proton (b) collisions and results
of simultaneous fits of these two parameters. The red line
corresponds to the fit at $\sqrt{s_{\footnotesize\mbox{min}}}=5$,
the blue line - at $\sqrt{s_{\footnotesize\mbox{min}}}=30$.
Experimental data are from \cite{pdg}.} \label{fig:fig-rho2}
\end{figure*}

The simultaneous fits were considered for various ensembles for
scattering parameters. Results of simultaneous fit for
$\{\rho^{pp},\rho^{\bar{p}p}\}$ by (\ref{rhopp}) and
(\ref{rhopbp}) with all free parameters are shown at Fig.
\ref{fig:fig-rho2}a and Fig. \ref{fig:fig-rho2}b for $pp$ and
$\bar{p}p$, respectively. Significant increasing of fitted points
allows us to study the approximation of $\rho$-parameter energy
dependence up to highest $\sqrt{s_{\footnotesize\mbox{min}}}=30$.
Numerical values of fit parameters are in the Table
\ref{fig:tabela7}. As seen fit qualities are statistical
acceptable at $\sqrt{s_{\footnotesize\mbox{min}}}\geq 3$.
Parameter errors are the same order or some smaller than that for
individual fits at corresponding
$\sqrt{s_{\footnotesize\mbox{min}}}$. Constraints
(\ref{constraints}) are valid at
$\sqrt{s_{\footnotesize\mbox{min}}}\geq 15$ for all parameters
within two standard deviations at least. Approximation curves show
smooth behavior at all $\sqrt{s_{\footnotesize\mbox{min}}}$. Some
decreasing of $\rho^{pp}$ is observed at high energies (Fig.
\ref{fig:fig-rho2}a). This changing of $\rho^{pp}$ at high
energies is most noticeable at
$\sqrt{s_{\footnotesize\mbox{min}}}=15$. Approximation curves are
very close for $\rho^{\bar{p}p}$ for $0.03 \geq \sqrt{s} \leq 1$
TeV at all $\sqrt{s_{\footnotesize\mbox{min}}}$ under considered.
Behaviour of fit curves for $\rho^{\bar{p}p}(s)$ depends on value
of $\sqrt{s_{\footnotesize\mbox{min}}}$. At
$\sqrt{s_{\footnotesize\mbox{min}}}=15$ and 30 corresponding
curves demonstrate almost constant value $\rho^{\bar{p}p}\sim
0.12$ in $\sqrt{s}>100$ GeV energy domain. But model curves show
decreasing at high energies $\sqrt{s}>1$ TeV at all another values
of low boundary.

\begin{figure*}[t!]
\centering{\includegraphics[width=17.0cm,height=17.0cm]{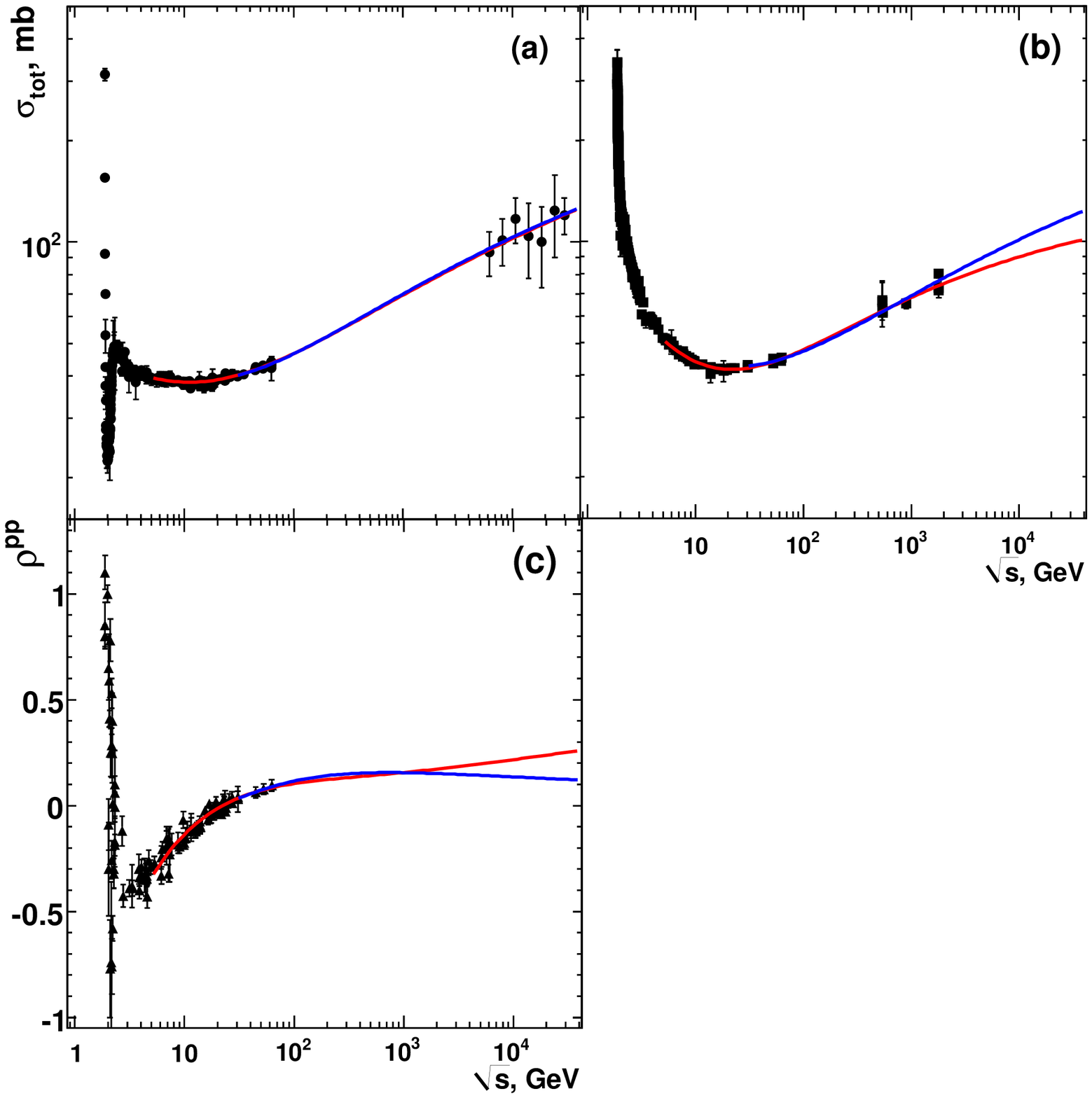}}
\caption{(color online) The $\sigma_{\mbox{\footnotesize{tot}}}$
energy dependence for proton-proton (a), antiproton-proton (b)
collisions, $\rho^{pp}$ parameter vs collision energy (c) and
results of simultaneous fits of these parameters. The red line
corresponds to the fit at $\sqrt{s_{\footnotesize\mbox{min}}}=5$,
the blue line - at $\sqrt{s_{\footnotesize\mbox{min}}}=30$.
Experimental data are from \cite{pdg}.}
\label{fig:fig-SigmaNNRhoPP}
\end{figure*}

Simultaneous fits for
$\{\sigma_{\mbox{\footnotesize{tot}}}^{pp},\sigma_{\mbox{\footnotesize{tot}}}^{\bar{p}p},\rho^{pp}
\}$ by (\ref{pp}), (\ref{pbp}) and (\ref{rhopp}) are shown at Fig.
\ref{fig:fig-SigmaNNRhoPP} for
$\sqrt{s_{\footnotesize\mbox{min}}}=5$ (red lines) and for
$\sqrt{s_{\footnotesize\mbox{min}}}=30$ (blue line). The values of
fit parameters are shown in Table \ref{fig:tabela8}. Suggested
parameterizations allow us to obtain statistically acceptable fit
qualities at $\sqrt{s_{\footnotesize\mbox{min}}}\geq5$ but fit
quality is quite reasonable even at lowest
$\sqrt{s_{\footnotesize\mbox{min}}}=3$. The precisions are
reasonable for all parameters. Constrains (\ref{constraints}) are
valid at $\sqrt{s_{\footnotesize\mbox{min}}}\geq 15$ for all
parameters (with exception of $\delta$, $\bar{\delta}$ at two
lowest $\sqrt{s_{\footnotesize\mbox{min}}}=15,~20$ under
considered in this case) within two standard deviations at least.
Total cross-section curves for $pp$ show a very close behavior
both for $\sqrt{s_{\footnotesize\mbox{min}}}=5$ and
$\sqrt{s_{\footnotesize\mbox{min}}}=30$. Model curve for
$\sigma_{\mbox{\footnotesize{tot}}}^{\bar{p}p}$ at
$\sqrt{s_{\footnotesize\mbox{min}}}=5$ shows significantly slow
increasing for high energy domain than that at
$\sqrt{s_{\footnotesize\mbox{min}}}=30$. The approximation curve
for $\rho^{pp}$ at $\sqrt{s_{\footnotesize\mbox{min}}}=5$
demonstrates the opposite behaviour at collision energy increasing
in high energy domain ($\sqrt{s} > 1$ TeV) than that the curve at
highest low boundary for fitted energy domain.

\begin{figure*}[t!]
\centering{\includegraphics[width=17.0cm,height=17.0cm]{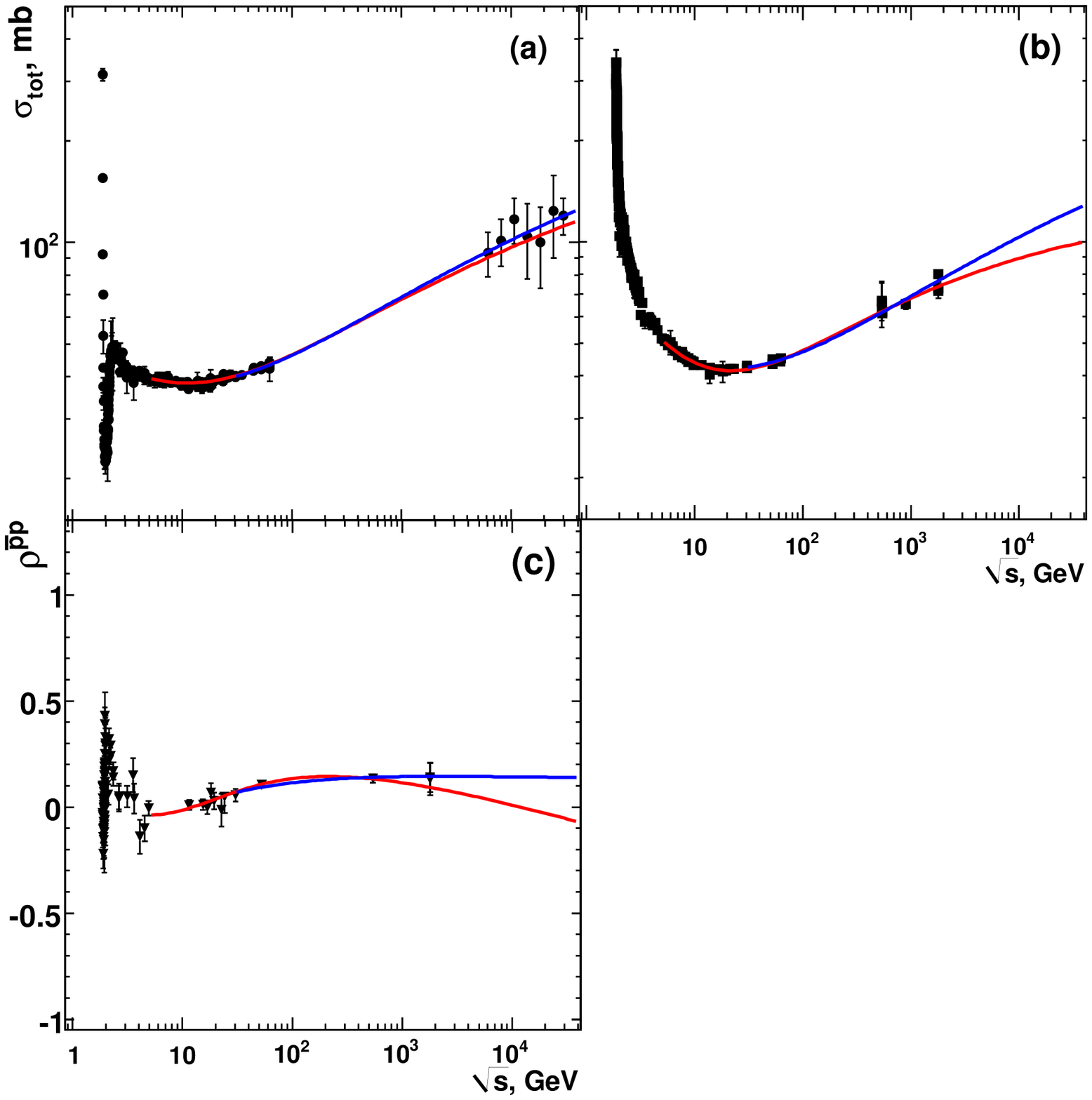}}
\caption{(color online) The $\sigma_{\mbox{\footnotesize{tot}}}$
energy dependence for proton-proton (a), antiproton-proton (b)
collisions, $\rho^{\bar{p}p}$ parameter vs collision energy (c)
and results of simultaneous fits of these parameters. The red line
corresponds to the fit at $\sqrt{s_{\footnotesize\mbox{min}}}=5$,
the blue line - at $\sqrt{s_{\footnotesize\mbox{min}}}=30$.
Experimental data are from \cite{pdg}.}
\label{fig:fig-SigmaNNRhoPPbar}
\end{figure*}

Results for simultaneous fits of
$\{\sigma_{\mbox{\footnotesize{tot}}}^{pp},
\sigma_{\mbox{\footnotesize{tot}}}^{\bar{p}p}, \rho^{\bar{p}p} \}$
by (\ref{pp}), (\ref{pbp}) and (\ref{rhopbp}) are shown at Fig.
\ref{fig:fig-SigmaNNRhoPPbar} and in Table \ref{fig:tabela9}.
Conclusions based on the numerical parameter values are similar
them which have described above for Table \ref{fig:tabela8}. One
can see that fit parameters have a close values both in Table
\ref{fig:tabela8} and in Table \ref{fig:tabela9} at the same value
of lower boundary for energy domain at
$\sqrt{s_{\footnotesize\mbox{min}}}\geq 3$ with exception of
$k$-parameter. The total cross-section for $\bar{p}p$ shows
significantly slow increasing in high energy domain at
$\sqrt{s_{\footnotesize\mbox{min}}}=5$ than that at
$\sqrt{s_{\footnotesize\mbox{min}}}=30$ for the ensemble of
scattering parameters under study (Fig.
\ref{fig:fig-SigmaNNRhoPPbar}b). Behaviour of model curves for
$\rho^{\bar{pp}}$ at high energies depends on value of low
boundary. The curves show noticeable decreasing at
$\sqrt{s_{\footnotesize\mbox{min}}}=5-15$ for energy range
$\sqrt{s} > 1$ TeV. But fit curves demonstrate almost constant
value $\rho^{\bar{p}p} \sim 0.10 - 0.15$ at higher values of low
boundary for energy domain indicated above.

\begin{figure*}[t!]
\centering{\includegraphics[width=17.0cm,height=17.0cm]{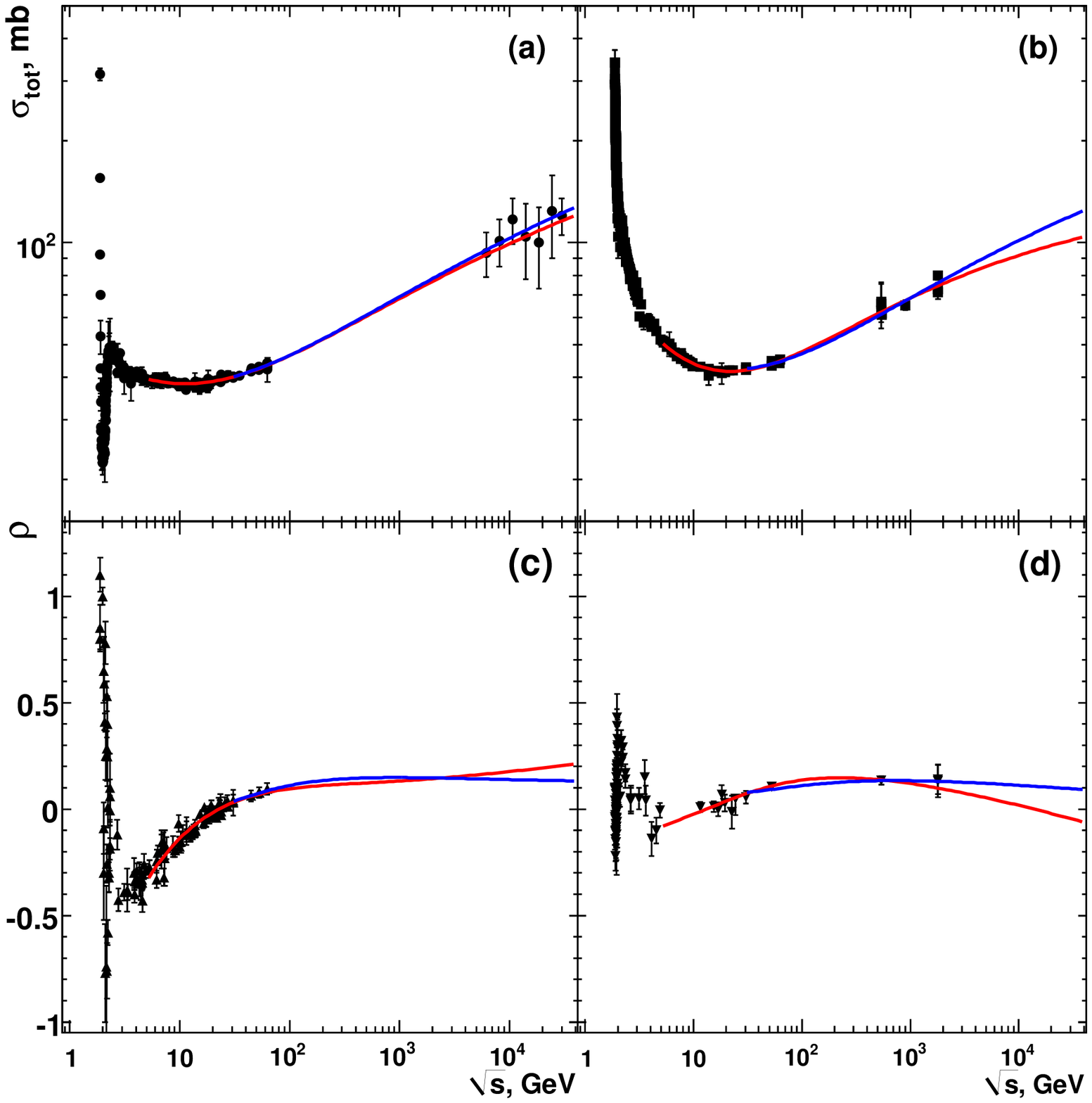}}
\caption{(color online) The $\sigma_{\mbox{\footnotesize{tot}}}$
energy dependence for proton-proton (a), antiproton-proton (b)
collisions, $\rho^{pp}$ (c), $\rho^{\bar{p}p}$ (d) parameters vs
collision energy and results of simultaneous fits of all four
parameters. The red line corresponds to the fit at
$\sqrt{s_{\mbox{\footnotesize{min}}}}=5$, the blue line - at
$\sqrt{s_{\mbox{\footnotesize{min}}}}=30$. Experimental data are
from \cite{pdg}.} \label{fig:fig-SigmaNNRhoNN}
\end{figure*}

Simultaneous fits of full set of global scattering parameters
$\left\{\sigma_{t}^{pp},
\sigma_{t}^{\bar{p}p},\rho^{pp},\rho^{\bar{p}p}\right\}$ by
(\ref{pp}), (\ref{pbp}), (\ref{rhopp}) and (\ref{rhopbp}) are
shown at Fig. \ref{fig:fig-SigmaNNRhoNN} at
$\sqrt{s_{\footnotesize\mbox{min}}}=5$ (red lines) and at
$\sqrt{s_{\footnotesize\mbox{min}}}=30$ (blue lines). Numerical
values of fit parameters are shown in Table \ref{fig:tabela10}.
Fit quality is statistically acceptable for
$\sqrt{s_{\footnotesize\mbox{min}}} \geq 5$. Moreover fit quality
is reasonable at lowest value of
$\sqrt{s_{\footnotesize\mbox{min}}}$ under study. Thus fit
functions (\ref{pp}), (\ref{pbp}), (\ref{rhopp}) and
(\ref{rhopbp}) allow us to obtain a reasonable description of
experimental energy dependencies of global scattering parameters
in wide energy domain. On the other hand one needs to emphasize
that noticeable increasing (decreasing) of $\rho^{pp}$
($\rho^{\bar{p}p}$) value is observed for high energies at
$\sqrt{s_{\footnotesize\mbox{min}}} \leq 10$ (Fig.
\ref{fig:fig-SigmaNNRhoNN}c,d) especially at lowest boundary value
$\sqrt{s_{\footnotesize\mbox{min}}}=3$. At larger
$\sqrt{s_{\footnotesize\mbox{min}}}$ there is a maximum in
$\rho(s)$ dependence with following much slower decreasing of
$\rho$ for energy domain $\sqrt{s} > 1$ TeV both in proton-proton
and antiproton-proton interactions. Precisions are reasonable for
all fit parameters at any low boundary values. These precisions
are the same order or some smaller for this case than that for
individual fits of scattering parameters. Constraints
(\ref{constraints}) are valid for all fit parameters at
$\sqrt{s_{\footnotesize\mbox{min}}} \geq 20$ within errors with
exception for $\delta$ and $\bar{\delta}$. The last two parameters
have similar values, which are approached to each other with
boundary value increasing. For $\delta$ and $\bar{\delta}$
parameters constraints (\ref{constraints}) are valid at
$\sqrt{s_{\footnotesize\mbox{min}}}$ indicated above within two
standard deviations. Parameters $\alpha$, $\bar{\alpha}$, $\beta$,
$\bar{\beta}$ are almost constants within errors for
$\sqrt{s_{\footnotesize\mbox{min}}} \geq 15$.

The expected asymptotic behaviors are obtained for approximation
curves for all scattering parameters under study at highest value
of low boundary at least (Fig. \ref{fig:fig-SigmaNNRhoNN}). Curves
for both $pp$ and $\bar{p}p$ cross-sections shows a close behavior
at all $\sqrt{s_{\footnotesize\mbox{min}}}$ especially for
proton-proton interactions. Corresponding curves for
$\sigma^{\bar{p}p}$ demonstrate some differences for energies
higher than 1 TeV at low $\sqrt{s_{\footnotesize\mbox{min}}} \leq
5$ only. Difference between curves for $\rho^{pp}
\left(\rho^{\bar{p}p}\right)$ at several
$\sqrt{s_{\footnotesize\mbox{min}}}$ are larger, especially for
$\sqrt{s} \geq 1$ TeV energy domain.

Predictions are obtained for
$\sigma_{\mbox{\footnotesize{tot}}}^{pp}$,
$\sigma_{\mbox{\footnotesize{tot}}}^{\bar{p}p}$ at various
energies based on the fit results of full set of global scattering
parameters from Table \ref{fig:tabela10}. Total cross-section
values are shown in the Table \ref{fig:tabela11}. These results
are in agreement with predictions above from Table
\ref{fig:tabela4}. Uncertainties for these predictions are the
same order or smaller for the case under considered than results,
which were obtained for individual fits (Table \ref{fig:tabela4}).
The smallest uncertainties are at
$\sqrt{s_{\footnotesize\mbox{min}}}=15$. Perhaps, the one of the
reasons for such feature is the optimal value of low boundary
$\sqrt{s_{\footnotesize\mbox{min}}}$ which allows to exclude the
influence of spread of experimental points at low energies on
uncertainties of fit parameters as well as to remain amounts of
experimental data large enough for reasonable precision of fit. As
seen from the Table \ref{fig:tabela11} results for various
$\sqrt{s_{\footnotesize\mbox{min}}}$ are in a good agreement
within errors.

Our results coincide well with values of
$\sigma_{\mbox{\footnotesize{tot}}}^{\bar{p}p}$ measured by UA4
and CDF Collaborations at $\sqrt{s}=546$ GeV at any
$\sqrt{s_{\footnotesize\mbox{min}}}$, in particular, at
$\sqrt{s_{\footnotesize\mbox{min}}}=15$ which correspond to the
best precision for predictions as well as statistically acceptable
fit quality (see Table \ref{fig:tabela10}). The similar situation
at higher energy $\sqrt{s}=1.8$ TeV taking into account the wider
dispersion of experimental values obtained by E710 and CDF
Collaborations. For $\sqrt{s_{\footnotesize\mbox{min}}}$ under
considered above our predictions are in good agreement with Fly's
Eye Collaboration \cite{cosmic-ray2} data at $\sqrt{s}=30$ TeV but
results in Table \ref{fig:tabela11} are in qualitative agreement
only with the estimation $\sigma_{\mbox{\footnotesize{tot}}}^{pp}$
at $\sqrt{s}=40$ TeV \cite{gaisser} with taking into account
errors. Total cross-section value at very high energy
$\sqrt{s}=500$ TeV significantly smaller than that from
\cite{blockcahn} as well as corresponding value obtained for
individual fit results.

It is important to stress that simultaneous fits are used because
at high energies $pp$ and $\bar{p}p$ elastic scattering amplitudes
are (practically) the same function. Then, $pp$ and $\bar{p}p$
experimental data tend to the same value at high energies and we
expect that the absence of $pp$ experimental data at some high
energy value may be filled by presence of $\bar{p}p$ experimental
data at (almost) same energy and vice-versa. Therefore, we expect
that $pp$ and $\bar{p}p$ simultaneous fit allows more reliable
numerical results to the fit parameters. As can be viewed in
Tables \ref{fig:tabela4} and \ref{fig:tabela10} the errors in the
parameters obtained from simultaneous fits are the same order or
some smaller than those from individual fits allowing a more
precise description of the fitted physical quantity.

Based on the above study one can conclude that suggested
approximations allow to obtain fits (both individual and
simultaneous) of experimental data for global scattering
parameters with good statistical qualities for wide energy domain
$\sqrt{s} \geq 5$ GeV. Moreover this energy domain can be expanded
as $\sqrt{s} \geq 3$ GeV in some cases. The quality of last
simultaneous fit at $\sqrt{s_{\footnotesize\mbox{min}}}=10$ is
better than that for Kang-Nicolescu model fit as well as for
Donnachie-Landshoff fit of $pp$ and $\bar{p}p$ experimental data
\cite{menon}. Fit qualities obtained in this work for wide energy
domains are close to the best fit qualities observed previously
for various process in the framework of model with different
number of poles
\cite{compete1,compete2,Cudell-PRD-61-034019-2000,Cudell-PRD-73-034008-2006}.

\section{Comparisons}\label{secY}
In this section, we compare the possible physical description of our model with some models present in the literature. There are
several models but for the sake of the simplicity, we restrict ourselves to three models only. The first one is the famous and
successful Donnachie-Landshoff model \cite{dl} based on a very simple parametrization structure evolving the exchange of
reggeons and pomerons trajectories. The second one is the Kharzeev-Levin model \cite{kl1,kl2,kl3} based on N=4
Super-symmetric Yang-Mills theory and considering the contribution of D-instantons for multiparticle production. Finally, the
Godbole-Grau-Pancheri-Srivastava model \cite{ggps1,ggps2,ggps3} based on minijet contributions to the total cross-section.
\subsection{Donnachie-Landshoff Model}\label{Y.1}
One of the most successful models is the Donnachie-Landshoff (DL) parametrization \cite{dl}
\begin{eqnarray} \label{dl}
\sigma_{\mbox{\footnotesize{tot}}}^{pp}(s)=Xs^{\epsilon}+Y^{pp}s^{-\eta},
\
\sigma_{\mbox{\footnotesize{tot}}}^{\bar{p}p}(s)=Xs^{\epsilon}+Y^{\bar{p}p}s^{-\eta},
\end{eqnarray}

\noindent where $X$, $Y^{pp}$, $Y^{\bar{p}p}$, $\epsilon$, and $\eta$ are fit parameters. The first term on the right hand side
of the above expressions ($X=21.7$ and $\epsilon\approx 0.08$) is associated with the pomeron contribution and has a common value to
both $pp$ and $\bar{p}p$ cross-section because the pomeron carries the quantum number of the vacuum and cannot distinguish particles
and antiparticles. The second term ($Y^{pp}\approx 56.1$, $Y^{\bar{p}p}\approx 98.4$ and $\eta\approx 0.45$) is associated
with the reggeon trajectory, and it may be different to particles and antiparticles at low $s$. This term may be identified as
resulting from $\rho$, $\omega$, $f_2$, $a_2$ exchange, for example. At $s\rightarrow\infty$, the difference between $pp$ and
$\bar{p}p$ total cross-section parametrization can increases at most as $\ln s$. However, the difference may vanish if $\sigma_{tot}^{pp}=\sigma_{tot}^{\bar{p}p}$ and, in any case, the ratio between both total cross sections tends to 1 as stated by the Pomeranchuk theorem.

The DL model possesses clearly two regimes. Roughly speaking, at low energy the second term at the right hand side of Eqs.
(\ref{dl}) is the leading one (the reggeon exchange dominates) and at high energy regime, the first term will be the leading one (the
pomeron exchange dominates). This picture is quite similar to our model where the first term of the right hand side of (\ref{pp})
and (\ref{pbp}) is the sub-leading one and may be associated to the exchange of a reggeon trajectory with intercept
$\alpha_R(0)\approx 0.95$.

On the other hand, using (\ref{pomeron}) we may connect (at high energies) the behavior of the leading term of our model with the
exchange of a pomeron trajectory. Adopting $c=21.7$ as obtained in the DL model we obtain $p\leq 0.06$ and therefore,
$\alpha_P(0)\leq 1.06$. This value is slightly above 1 and may leads to the violation of the Froissart-Martin bound, but it does
not occur for momenta lower than Planck scale. In this way, the pomeron exchange mechanism may ensures the preservation of
unitarity.

However, comparisons done above are not exactly correct. In our model, at low energies, reggeon and pomeron trajectories interacts
with each other mixing the intercept contributions. Therefore, the reggeon intercepts showed here present pomeron contributions.
Indeed, $f_2$ trajectory, for example, has the quantum number of the vacuum as the pomeron. There may be contributions from
pomerons at low energies, preventing the complete separation between pomerons and reggeons. We expect, at high energies, the
divergence of reggeon and pomeron trajectories from each other. Then, the influence of reggeons will ceases and only the pomeron
exchange will be the main physical mechanism restoring the validity of Pomeranchuk theorem.

\subsection{Kharzeev-Levin Model}\label{Y.2}
The Kharzeev-Levin (KL) model \cite{kl1,kl2,kl3} is based on the Super-symmetric Yang-Mills theory in N=4 (N=4 SYM). Considering
contributions of D-instantons in $AdS_5$ bulk space, they argue that D-instantons coupled to dilatons and axions are responsible
for multiparticle production in strongly coupled N=4 SYM. As a consequence, the cross-section increases with the energy
\begin{eqnarray}
\nonumber \sigma_{\mbox{\footnotesize{inel}}}\propto s^{\Delta-2},
\end{eqnarray}

\noindent where $\Delta$ depends on the convenient choice of parameters \cite{kl1,kl2,kl3}. This approach is interesting since
it suggests that topological effects may be important in high energy collisions. Furthermore, they argue that the weakly coupled
graviton generates the elastic amplitude and the correspondent part of the total cross-section. On the other hand, D-instanton
induced interactions of dilatons and axions are responsible for the multiparticle production processes.

This formalism predicts that the cross-section will increase with the energy due to D-instanton mechanism. This mechanism is induced
by interactions of dilatons and axions responsible for the multiparticle production process. The pomeron intercepts predicted
in KL model depends on the value of $\Delta$ (and $\sigma_{\mbox{\footnotesize{inel}}}\propto s^{\Delta-2}$). They
found $\Delta\approx 4$ and consequently, the slope $\alpha'_{P}\approx 0.5$ GeV$^{-2}$.

However, this value of $\Delta$ may imply in a strong unitarisation scheme to restore the Froissart-Martin bound. As
shown above, the pomeron exchange may be the main mechanism responsible for restore the unitarisation. Indeed, they obtain a
very high pomeron intercept, possibly indicating the exchange of a new soft pomeron trajectory family ($\Delta$ is not unique).

The model proposed here possesses a very smooth increase (at high energies) when compared with the standard theoretical result
$\sigma_{\mbox{\footnotesize{tot}}}\approx 60\ln^2s$. The KL model may explains these differences in the behaviors by the mechanism
of multiparticle production: at high energies the multiparticle mechanism acts, increasing the hadron cross-section and when the
saturation limit approaches the pomeron exchange has become physically important restoring the unitarity scheme turning the
increase of the total cross-section smooth as obtained. Moreover, we may suppose a mixed state "reggeon+pomeron" as being the responsible for the observed differences between $pp$ and $\bar{p}p$ total cross-section.

\subsection{Godbole-Grau-Pancheri-Srivastava Model}\label{Y.3}
The study of QCD minijet contribution is the main goal of the Godbole-Grau-Pancheri-Srivastava (GGPS) model
\cite{ggps1,ggps2,ggps3}. This model includes a re-summation of soft gluon radiation which they argue it is necessary to tame the
fast high energy rise typical of a purely perturbative minijet model.

The minijet formalism predicts a very fast rise to $\sigma_{\mbox{\footnotesize{jet}}}\approx s^{0.3}$ at high
energies. To restore the total cross-section they suppose that soft gluon emissions from the colliding partons are the physical
attenuation mechanism. To restore the Froissart-Martin bound, they obtain a "new" formulation of the Froissart-Martin bound
\begin{eqnarray} \nonumber
\sigma_{\mbox{\footnotesize{tot}}}\sim (\epsilon\ln s)^{1/n}
\end{eqnarray}

\noindent where $\epsilon$ fixes the asymptotic rise and $1/2<n<1$
modulates the infrared behavior of $\alpha_s$
\cite{ggps1,ggps2,ggps3}.

Pomeron exchange is believed to be the exchange of a system of gluons. Therefore (ultimately), the pomeron exchange will be the
Froissart-Martin bound restoration mechanism as in the KL model. As seen above, the pomeron exchange has a very important role if
we want to understand the taming mechanism behind no saturation effect in our model.

\section{\label{sec7}Final Remarks}

The model proposed here to describe the hadron-hadron total cross-section and the $\rho$ parameter is basically based on two
rigorously bounds obtained in the framework of AQFT. Namely, the Froissart-Martin bound and Jin-Martin-Cornille bound. Roughly
speaking, in this model the like-Froissart-Martin bound will control the high energy behavior of the total cross-section, and like-Jin-Martin-Cornille bound will defines its behavior at low energies. However, the analysis of our model is not so simple,
since we suppose an attenuation mechanism in the Froissart-Martin bound $\sim s^{-\alpha}\ln^2s$ which mix these two bounds,
especially at low energies.

The statistical description of $\sigma_{\mbox{\footnotesize{tot}}}$ and $\rho$ experimental data
is acceptable. The quality of some simultaneous fits is better than that for Kang-Nicolescu model fit as well as for
Donnachie-Landshoff fit of $pp$ and $\bar{p}p$ experimental data. The qualities of the fits obtained in this work for wide energy
domains are closer to the best fit qualities previously observed for several processes in the framework of the model with a
different number of poles. Therefore, based on this statistical fact, it is possible perform predictions in the energy variable
and the results can be compared with others present in some models and parameterizations in available literature. Results viewed here
shown some agreement among themselves.

Comparisons with some models may allow some physical understanding of what is happening here. In the context of DL model, we may
understand the the index in the low energy term of our model as representing a reggeon exchange trajectory, and the index in the high energy term - as a pomeron exchange trajectory. Yet, the picture viewed here seems to be not so simple. The mixed description proposed, especially at low energies, prevents the identification of each index contribution separately (or each trajectory separately) and, therefore, we cannot treat this low energy term as being a pure reggeon exchange. 


In the context of KL and GGPS models, the pomeron exchange is the responsible for the attenuation of the rise of the total
cross-section. However, in both models the saturation of the total cross-section is quickly reached and therefore, it is necessary
the exchange of not only one pomeron but a whole family. In our model, this attenuation mechanism may be explained by the
interaction of pomerons and reggeons that tames the rise of the total cross-section and this mixed state of index (trajectories) may be responsible for the observed differences between $pp$ and $\bar{p}p$ behaviors.

On the other hand, the model presented here does not saturate the
total cross-section. Then, we may suppose that this is due to the
exchange of more than only one pomeron trajectory since the
increase of experimental data set in the fit procedures may give
rise to different values of the effective index $p$ and each value
may represent a novel pomeron trajectory. Therefore, novel
experimental results to total cross section can be added to the
original data set allowing more precise values to the effective
index $p$ through fit procedures. These trajectories may tend to a
limiting value, i.e, $p_i\rightarrow p_l$ where $i$ represents a
pomeron trajectory and $l$ represents the limiting trajectory
($\alpha_{P_i}(0)\rightarrow \alpha_{P_l}(0)$). If this pomeron
trajectory limit exist it may represent a definite (or not, since
a new physical attenuation mechanism may be discovered) bound to
the rise of total cross section. This behavior at very high
energies is one of the puzzles to be solved at the future HC
because the applied mechanisms used to tame its very fast rise, we
hope, will be verified.

\section*{Acknowledgments}
S.D.C is grateful to UFSCar for the financial support. The work of V.A.O. was supported partly by the Russian Federal Agency
for Science and Innovation grant (State Contract No. 02.740.11.2040).


\clearpage
\begin{table*}[ht]
\begin{center}
\begin{tabular}{c|ccccc}
\noalign{\smallskip}\hline\noalign{\smallskip}
$\sqrt{s_{\footnotesize\mbox{min}}}$ & $\alpha$ &  $\beta \times 10^{3}$ & $-\gamma$ & $\delta$ & $\chi^2/\mathrm{n.d.f}$  \\
\hline
3                      &0.028$\pm$0.008 &3.9$\pm$0.3 & 5.882$\pm$0.005& 51.8$\pm$0.5 &107.6/139        \\
5                      &0.020$\pm$0.010 &3.4$\pm$0.5 & 5.890$\pm$0.010& 50.8$\pm$1.5 &74.9/107           \\
10                     &0.010$\pm$0.012 &2.9$\pm$0.6 & 5.910$\pm$0.020& 48.5$\pm$2.8 &46.7/69            \\
15                     &0.010$\pm$0.008 &2.9$\pm$0.3 & 5.906$\pm$0.005& 48.8$\pm$1.0 &34.2/58            \\
20                     &0.019$\pm$0.003 &3.6$\pm$0.1 & 5.876$\pm$0.004& 54.8$\pm$1.1 &30.1/45           \\
25                     &0.031$\pm$0.002 &4.6$\pm$0.1 & 5.821$\pm$0.009& 69.5$\pm$4.2 &15.9/33               \\
30                     &0.021$\pm$0.012 &3.7$\pm$0.9 & 5.870$\pm$0.040& 55.7$\pm$10.1&15.1/30             \\
\noalign{\smallskip} \hline
\end{tabular}
\caption{Parameters for fitting by (\ref{pp}) of
$\sigma_{\mbox{\footnotesize{tot}}}^{pp}$ energy
dependence.}\label{fig:tabela1}
\end{center}
\vspace*{0.5cm}
\begin{center}
\begin{tabular}{c|ccccc}
\noalign{\smallskip}\hline\noalign{\smallskip}
$\sqrt{s_{\footnotesize\mbox{min}}}$ & $\bar{\alpha}$ &  $\bar{\beta} \times 10^{3}$ & $-\bar{\gamma}$ & $\bar{\delta}$ & $\chi^2/\mathrm{n.d.f}$  \\
\hline
3                      &0.085$\pm$0.008 &9.7$\pm$0.3 & 5.664$\pm$0.006& 130.9$\pm$1.8 &160.6/69                           \\
5                      &0.059$\pm$0.006 &6.9$\pm$0.5 & 5.749$\pm$0.013&  99.8$\pm$3.8 &35.7/55                             \\
10                     &0.040$\pm$0.013 &5.0$\pm$1.0 & 5.808$\pm$0.034&  80.8$\pm$9.0 &22.4/25                               \\
15                     &0.005$\pm$0.026 &2.7$\pm$1.2 & 5.895$\pm$0.043&  58.1$\pm$8.5 &16.3/20                           \\
20                     &0.013$\pm$0.032 &3.1$\pm$1.8 & 5.875$\pm$0.067&  63.6$\pm$14.8&15.2/13                             \\
25                     &0.023$\pm$0.031 &3.8$\pm$2.1 & 5.836$\pm$0.094&  77.1$\pm$28.1&14.8/11                               \\
\noalign{\smallskip} \hline
\end{tabular}
\caption{Parameters for fitting by (\ref{pbp}) of
$\sigma_{\mbox{\footnotesize{tot}}}^{\bar{p}p}$ energy
dependence.}\label{fig:tabela2}
\end{center}
\vspace*{0.5cm}
\begin{center}
\begin{tabular}{c|cccccccc}
\noalign{\smallskip}\hline\noalign{\smallskip}
          &                &                &                &$\sqrt{s_{\footnotesize\mbox{min}}}$ &   &    &                  \\
$\sqrt{s}$(TeV) &3        &5               &10              &15              &20             &25              &30            \\
\hline
          &                &                &                &$\sigma_{\mbox{\footnotesize{tot}}}^{pp}$ (mb)        &                & & &           \\
0.2       &51.95 $\pm$4.32 &51.06$\pm$6.62  &52.09$\pm$9.09  &51.42 $\pm$4.58 &52.42$\pm$1.73  &52.84$\pm$1.74  &52.0$\pm$12.3 \\
0.5       &60.46 $\pm$6.14 &59.66$\pm$9.15  &61.0$\pm$12.2   &60.29 $\pm$6.55 &61.82$\pm$2.41  &62.64$\pm$2.06  &61.3$\pm$15.5 \\
7         &91.4 $\pm$13.7  &92.6 $\pm$19.7  &96.7 $\pm$26.0  &96.1 $\pm$15.3  &98.04$\pm$5.53  &97.92 $\pm$3.98 &96.8 $\pm$29.7 \\
10        &96.0 $\pm$15.0  &97.7 $\pm$21.5  &102.6$\pm$28.5  &102.0$\pm$16.8  &103.75$\pm$6.09 &103.13$\pm$4.33 &102.3$\pm$32.2 \\
14        &100.4$\pm$16.3  &102.7$\pm$23.3  &108.3$\pm$30.9  &107.7$\pm$18.4  &109.26$\pm$6.64 &108.09$\pm$4.67 &107.7$\pm$34.6 \\
30        &110.7$\pm$19.3  &114.4$\pm$27.7  &122.0$\pm$37.0  &121.4$\pm$22.3  &122.20$\pm$8.01 &119.42$\pm$5.50 &120.2$\pm$40.5 \\
40        &114.6$\pm$20.6  &118.9$\pm$29.4  &127.4$\pm$39.5  &126.8$\pm$23.9  &127.23$\pm$8.56 &123.71$\pm$5.83 &125.0$\pm$42.9 \\
50        &117.7$\pm$21.5  &122.4$\pm$30.8  &131.6$\pm$41.5  &131.1$\pm$25.2  &131.18$\pm$9.00 &127.04$\pm$6.09 &128.8$\pm$44.7 \\
100       &127.2$\pm$24.7  &133.7$\pm$35.5  &145.4$\pm$48.2  &144.9$\pm$29.4  &143.7$\pm$10.5  &137.35$\pm$6.94 &140.7$\pm$50.9 \\
200       &136.7$\pm$28.1  &145.3$\pm$40.4  &159.8$\pm$55.4  &159.3$\pm$34.1  &156.5$\pm$12.1  &147.56$\pm$7.84 &152.9$\pm$57.4 \\
500       &149.3$\pm$32.9  &160.8$\pm$47.5  &179.8$\pm$66.1  &179.4$\pm$41.0  &173.9$\pm$14.3  &160.79$\pm$9.10 &169.3$\pm$66.7 \\
\hline
          &                &                &                &$\sigma_{\mbox{\footnotesize{tot}}}^{\bar{p}p}$ (mb)        &               & & &           \\
0.546     &62.10$\pm$2.94&61.93$\pm$6.17&61.9$\pm$14.0&61.4$\pm$26.8&60.8$\pm$35.6&61.3$\pm$36.9&- \\
1.8       &70.37$\pm$3.80&73.22$\pm$8.30&74.9$\pm$19.8&76.2$\pm$38.8&75.1$\pm$50.3&75.6$\pm$50.7&- \\
\noalign{\smallskip} \hline
\end{tabular}
\caption{Predictions for $\sigma_{\mbox{\footnotesize{tot}}}$ at
various energies based on the individual fits of energy dependence
of proton-proton and proton-antiproton total
cross-sections.}\label{fig:tabela3}
\end{center}
\vspace*{0.5cm}
\begin{center}
\begin{tabular}{c|ccccc}
\noalign{\smallskip}\hline\noalign{\smallskip}
$\sqrt{s_{\footnotesize\mbox{min}}}$ & $\alpha_{NN}$ & $\beta_{NN} \times 10^{3}$ & $-\gamma_{NN}$ & $\delta_{NN}$ & $\chi^2/\mathrm{n.d.f}$  \\
\hline
15                     &0.042$\pm$0.008 &5.3$\pm$0.7 & 5.804$\pm$0.028&  73.7$\pm$7.7 &519.3/82                               \\
20                     &0.034$\pm$0.011 &4.6$\pm$0.9 & 5.837$\pm$0.036&  64.1$\pm$8.8 &196.9/62                           \\
25                     &0.011$\pm$0.017 &2.9$\pm$1.0 & 5.912$\pm$0.040&  47.3$\pm$7.0 &91.1/48                             \\
30                     &0.010$\pm$0.003 &2.8$\pm$0.1 & 5.915$\pm$0.004&  46.6$\pm$1.1 &90.8/45                               \\
\noalign{\smallskip} \hline
\end{tabular}
\caption{Parameters for fitting by (\ref{pp}) of
$\sigma_{\mbox{\footnotesize{tot}}}^{NN}$ energy
dependence.}\label{fig:tabela4}
\end{center}
\end{table*}

\begin{table*}
\begin{center}
\begin{tabular}{lccccccc}
\hline \multicolumn{1}{l}{Parameter} &
\multicolumn{7}{c}{$\sqrt{s_{\footnotesize\mbox{min}}}$} \\
\cline{2-8} \rule{0pt}{10pt}
 & 3 & 5 & 10 & 15 & 20 & 25 & 30 \\
\hline
 &   &   &    & i) &    &    &    \\
\hline
$k$, mbarn                 &$39 \pm 6$      &$-0.5 \pm 17$   &$-239 \pm 36$     &$15 \pm 60$       &$-115 \pm 104$ &$571 \pm 239$ &$648 \pm 301$ \\
$\chi^{2}/\mbox{n.d.f.}$   &$146/80$        &$93.9/63$       &$39.1/47$         &$32.3/33$         &$13.8/22$      &$1.53/9$      &$0.16/6$      \\
\hline
 &   &   &    & ii) &    &    &    \\
\hline
$k$, mbarn                 &$32 \pm 8$      &$150 \pm 19$    &$514 \pm 2$       &$1202 \pm 2$      &$40.0 \pm 1.4$      & - & - \\
$\alpha$                   &$0.54 \pm 0.05$ &$0.55 \pm 0.09$ &$0.53 \pm 0.06$   &$-0.20 \pm 0.11$  &$-0.15 \pm 0.12$    & - & - \\
$\beta \times 10^{3}$      &$4.4 \pm 0.5$   &$6 \pm 5$       &$0.8 \pm 0.2$     &$0.006 \pm 0.003$ &$0.0011 \pm 0.0002$ & - & - \\
$-\gamma$                  &$5.31 \pm 0.05$ &$5.21 \pm 0.10$ &$5.21 \pm 0.06$   &$4.4 \pm 0.1$     &$4.10 \pm 0.18$     & - & - \\
$\delta$, mbarn            &$40 \pm 20$     &$53 \pm 62$     &$50 \pm 27$       &$141 \pm 55$      &$129 \pm 12$        & - & - \\
$\bar{\alpha}$             &$0.59 \pm 0.17$ &$0.7 \pm 0.3$   &$0.7 \pm 0.5$     &$-0.18 \pm 0.07$  &$-0.13 \pm 0.11$    & - & - \\
$\bar{\beta} \times 10^{3}$&$0.66 \pm 0.07$ &$2 \pm 2$       &$0.057 \pm 0.005$ &$0.01 \pm 0.01$   &$0.0022 \pm 0.0003$ & - & - \\
$-\bar{\gamma}$            &$5.12 \pm 0.08$ &$5.04 \pm 0.07$ &$4.963 \pm 0.003$ &$5.010 \pm 0.006$ &$5.041 \pm 0.013$   & - & - \\
$\bar{\delta}$, mbarn      &$51 \pm 3$      &$181 \pm 38$    &$733 \pm 6$       &$1443 \pm 22$     &$40 \pm 3$          & - & - \\
$\chi^{2}/\mbox{n.d.f.}$   &$75.3/72$       &$63.9/55$       &$38.9/39$         &$26.0/25$         &$12.4/14$           & - & - \\
\hline
\end{tabular}
\caption{Parameters for individual fitting by (\ref{rhopp}) of
$\rho^{pp}$ energy dependence. Part i) corresponds to approach
with one free parameter $k$ and part ii) shows results for
approach with all free parameters in (\ref{rhopp}).}
\label{fig:tabela5}
\end{center}
\end{table*}

\begin{table*}
\begin{center}
\begin{tabular}{lccccccc}
\hline \multicolumn{1}{l}{Parameter} &
\multicolumn{7}{c}{$\sqrt{s_{\footnotesize\mbox{min}}}$} \\
\cline{2-8} \rule{0pt}{10pt}
 & 3 & 5 & 10 & 15 & 20 & 25 & 30 \\
\hline
 &   &   &    & i) &    &    &    \\
\hline
$k$, mbarn                 &$109 \pm 25$    &$65 \pm 144$      &$38 \pm 131$ &$-436 \pm 172$   &$-87 \pm 416$ &$1172 \pm 1197$ &$884 \pm 1197$ \\
$\chi^{2}/\mbox{n.d.f.}$   &$44.4/17$       &$20.8/11$         &$16.6/11$    &$2.73/10$        &$8.21/6$      &$11.7/4$        &$7.60/4$      \\
\hline
 &   &   &    & ii) &    &    &    \\
\hline
$k$, mbarn                 &$-0.3 \pm 0.2$  &$15 \pm 3$        & -           &$18 \pm 3$       & - & - & -\\
$\alpha$                   &$-0.4 \pm 0.2$  &$0.078 \pm 0.016$ & -           &$-0.03 \pm 0.10$ & - & - & -\\
$\beta \times 10^{6}$      &$0.13 \pm 0.10$ &$4.5 \pm 0.9$     & -           &$3 \pm 3$        & - & - & -\\
$-\gamma$                  &$4.0 \pm 1.3$   &$4.8 \pm 0.4$     & -           &$4.55 \pm 0.17$  & - & - & -\\
$\delta$, mbarn            &$-5 \pm 4$      &$10 \pm 11$       & -           &$61.5 \pm 48.2$  & - & - & - \\
$\bar{\alpha}$             &$-0.4 \pm 0.2$  &$0.056 \pm 0.015$ & -           &$-0.05 \pm 0.10$ & - & - & - \\
$\bar{\beta} \times 10^{6}$&$0.06 \pm 0.05$ &$3.4 \pm 0.6$     & -           &$2.0 \pm 1.3$    & - & - & - \\
$-\bar{\gamma}$            &$5.52 \pm 0.02$ &$5.07 \pm 0.06$   & -           &$5.12 \pm 0.12$  & - & - & - \\
$\bar{\delta}$, mbarn      &$0.5 \pm 0.5$   &$13 \pm 3$        & -           &$13 \pm 10$      & - & - & - \\
$\chi^{2}/\mbox{n.d.f.}$   &$9.67/9$        &$1.84/3$          & -           &$1.78/2$         & - & - & - \\
\hline
\end{tabular}
\caption{Parameters for individual fitting by (\ref{rhopbp}) of
$\rho^{\bar{p}p}$ energy dependence. Part i) corresponds to
approach with one free parameter $k$ and part ii) shows results
for approach with all free parameters in (\ref{rhopbp}).}
\label{fig:tabela6}
\end{center}
\end{table*}

\begin{table*}
\begin{center}
\begin{tabular}{l|ccccccc}
\hline \multicolumn{1}{l|}{Parameter} &
\multicolumn{7}{c}{$\sqrt{s_{\footnotesize\mbox{min}}}$} \\
 & 3 & 5 & 10 & 15 & 20 & 25 & 30 \\
\hline
$k$, mbarn                 & $11.7 \pm 3.5$    & $21 \pm 7$        & $-23 \pm 19$      & $109 \pm 42$      & $-242 \pm 68$     & $-174 \pm 147$    & $-200 \pm 189$   \\
$\alpha$                   & $0.05 \pm 0.02$   & $0.051 \pm 0.007$ & $0.038 \pm 0.006$ & $0.06 \pm 0.03$   & $0.045 \pm 0.006$ & $0.027 \pm 0.008$ & $0.05 \pm 0.03$ \\
$\beta \times 10^{3}$      & $0.92 \pm 0.18$   & $0.78 \pm 0.03$   & $1.12 \pm 0.06$   & $1.10 \pm 0.15$   & $1.06 \pm 0.07$   & $1.05 \pm 0.09$   & $1.3 \pm 0.7$     \\
$-\gamma$                  & $5.84 \pm 0.02$   & $5.844 \pm 0.016$ & $5.866 \pm 0.009$ & $5.68 \pm 0.06$   & $5.876 \pm 0.013$ & $5.84 \pm 0.02$   & $5.80 \pm 0.13$ \\
$\delta$, mbarn            & $10.05 \pm 0.49$  & $8.2 \pm 0.8$     & $13.3 \pm 0.8$    & $22 \pm 3$        & $10.6 \pm 1.0$    & $16 \pm 2$        & $18 \pm 12$    \\
$\bar{\alpha}$             & $0.061 \pm 0.005$ & $0.065 \pm 0.006$ & $0.043 \pm 0.005$ & $0.05 \pm 0.02$   & $0.045 \pm 0.006$ & $0.023 \pm 0.008$ & $0.03 \pm 0.02$ \\
$\bar{\beta} \times 10^{3}$& $1.18 \pm 0.10$   & $1.05 \pm 0.05$   & $1.28 \pm 0.08$   & $0.91 \pm 0.09$   & $1.11 \pm 0.09$   & $0.94 \pm 0.11$   & $0.9 \pm 0.3$     \\
$-\bar{\gamma}$            & $5.73 \pm 0.02$   & $5.705 \pm 0.004$ & $5.802 \pm 0.009$ & $5.75 \pm 0.05$   & $5.855 \pm 0.013$ & $5.887 \pm 0.016$ & $5.90 \pm 0.10$ \\
$\bar{\delta}$, mbarn      & $19.7 \pm 2.5$    & $18.7 \pm 0.6$    & $20.6 \pm 1.1$    & $16.2 \pm 1.9$    & $12.2 \pm 1.1$    & $12.9 \pm 1.5$    & $11 \pm 6$    \\
$\chi^{2}/\mbox{n.d.f.}$   & $86.6/90$         & $66.3/67$         & $42.1/51$         & $29.8/36$         & $12.8/21$         & $1.18/6$          & $0.25/3$          \\
\hline
\end{tabular}
\caption{Parameters for simultaneous fitting by (\ref{rhopp}) and
(\ref{rhopbp}) of $\rho$ energy dependencies for proton-proton and
antiproton-proton collisions respectively.} \label{fig:tabela7}
\end{center}
\end{table*}

\begin{table*}
\begin{center}
\begin{tabular}{l|ccccccc}
\hline \multicolumn{1}{l|}{Parameter} &
\multicolumn{7}{c}{$\sqrt{s_{\footnotesize\mbox{min}}}$} \\
 & 3 & 5 & 10 & 15 & 20 & 25 & 30 \\
\hline
$k$, mbarn                 & $42.0 \pm 6.5$    & $-52 \pm 18$      & $-131 \pm 42$     & $32 \pm 20$       & $107.6 \pm 74.5$  & $671.4 \pm 426.5$ & $621 \pm 454$     \\
$\alpha$                   & $0.016 \pm 0.007$ & $0.018 \pm 0.003$ & $0.011 \pm 0.004$ & $0.017 \pm 0.005$ & $0.025 \pm 0.010$ & $0.024 \pm 0.015$ & $0.022 \pm 0.008$ \\
$\beta \times 10^{3}$      & $3.4 \pm 0.3$     & $3.44 \pm 0.07$   & $3.0 \pm 0.1$     & $3.42 \pm 0.15$   & $4.1 \pm 0.6$     & $4.0 \pm 1.3$     & $3.8 \pm 0.6$     \\
$-\gamma$                  & $5.888 \pm 0.004$ & $5.889 \pm 0.001$ & $5.903 \pm 0.004$ & $5.885 \pm 0.003$ & $5.85 \pm 0.02$   & $5.85 \pm 0.07$   & $5.865 \pm 0.031$   \\
$\delta$, mbarn            & $51.3 \pm 0.4$    & $51.2 \pm 0.4$    & $49.1 \pm 0.9$    & $52.6 \pm 1.1$    & $60 \pm 6$        & $61 \pm 17$       & $57 \pm 8$        \\
$\bar{\alpha}$             & $0.084 \pm 0.003$ & $0.056 \pm 0.005$ & $0.042 \pm 0.011$ & $0.013 \pm 0.003$ & $0.026 \pm 0.018$ & $0.023 \pm 0.012$ & $0.023 \pm 0.011$ \\
$\bar{\beta} \times 10^{3}$& $9.6 \pm 0.3$     & $6.4 \pm 0.5$     & $5.15 \pm 0.90$   & $3.13 \pm 0.13$   & $4.0 \pm 1.3$     & $4 \pm 2$         & $4 \pm 2$     \\
$-\bar{\gamma}$            & $5.669 \pm 0.005$ & $5.758 \pm 0.012$ & $5.80 \pm 0.03$   & $5.879 \pm 0.002$ & $5.84 \pm 0.05$   & $5.84 \pm 0.10$   & $5.84 \pm 0.09$   \\
$\bar{\delta}$, mbarn      & $129.3 \pm 1.7$   & $97 \pm 3$        & $83 \pm 9$        & $61.8 \pm 1.2$    & $74 \pm 14$       & $77 \pm 34$       & $77 \pm 31$       \\
$\chi^{2}/\mbox{n.d.f.}$   & $392/288$         & $182/225$         & $109/141$         & $81.1/111$        & $60.2/80$         & $31.7/53$         & $30.1/47$         \\
\hline
\end{tabular}
\caption{Parameters for simultaneous fitting by (\ref{pp}),
(\ref{pbp}) and (\ref{rhopp}) of
$\sigma_{\mbox{\footnotesize{tot}}}^{pp}$,
$\sigma_{\mbox{\footnotesize{tot}}}^{\bar{p}p}$ and $\rho^{pp}$
energy dependencies.} \label{fig:tabela8}
\end{center}
\begin{center}
\begin{tabular}{l|ccccccc}
\hline \multicolumn{1}{l|}{Parameter} &
\multicolumn{7}{c}{$\sqrt{s_{\footnotesize\mbox{min}}}$} \\
 & 3 & 5 & 10 & 15 & 20 & 25 & 30 \\
\hline
$k$, mbarn                 & $102 \pm 20$      & $21 \pm 17$       & $-91 \pm 39$      & $-399 \pm 188$    & $-585 \pm 208$    & $-230 \pm 123$    & $-270.5 \pm 109.7$   \\
$\alpha$                   & $0.035 \pm 0.006$ & $0.026 \pm 0.004$ & $0.011 \pm 0.003$ & $0.018 \pm 0.004$ & $0.017 \pm 0.005$ & $0.018 \pm 0.004$ & $0.017 \pm 0.003$ \\
$\beta \times 10^{3}$      & $4.2 \pm 0.3$     & $3.77 \pm 0.13$   & $2.95 \pm 0.07$   & $3.38 \pm 0.11$   & $3.40 \pm 0.17$   & $3.31 \pm 0.15$   & $3.35 \pm 0.08$   \\
$-\gamma$                  & $5.878 \pm 0.005$ & $5.884 \pm 0.002$ & $5.906 \pm 0.002$ & $5.899 \pm 0.003$ & $5.884 \pm 0.002$ & $5.880 \pm 0.004$ & $5.889 \pm 0.005$ \\
$\delta$, mbarn            & $52.15 \pm 0.44$  & $51.6 \pm 0.4$    & $48.6 \pm 0.7$    & $51.5 \pm 1.1$    & $53.0 \pm 1.3$    & $54.1 \pm 1.2$    & $51.6 \pm 1.8$    \\
$\bar{\alpha}$             & $0.082 \pm 0.003$ & $0.058 \pm 0.002$ & $0.036 \pm 0.002$ & $0.014 \pm 0.003$ & $0.012 \pm 0.003$ & $0.016 \pm 0.013$ & $0.013 \pm 0.005$ \\
$\bar{\beta} \times 10^{3}$& $9.4 \pm 0.3$     & $6.63 \pm 0.12$   & $4.69 \pm 0.13$   & $3.15 \pm 0.11$   & $3.09 \pm 0.11$   & $3.3 \pm 0.7$     & $3.2 \pm 0.2$   \\
$-\bar{\gamma}$            & $5.669 \pm 0.005$ & $5.750 \pm 0.002$ & $5.818 \pm 0.003$ & $5.880 \pm 0.003$ & $5.878 \pm 0.006$ & $5.865 \pm 0.025$ & $5.870 \pm 0.007$ \\
$\bar{\delta}$, mbarn      & $129.7 \pm 1.7$   & $99.6 \pm 0.8$    & $78.1 \pm 1.3$    & $61.3 \pm 1.4$    & $63 \pm 3$        & $67 \pm 7$        & $65 \pm 5$        \\
$\chi^{2}/\mbox{n.d.f.}$   & $288/225$         & $115/173$         & $72.3/105$        & $53.4/88$         & $46.6/64$         & $31.7/48$         & $30.9/45$         \\
\hline
\end{tabular}
\caption{Parameters for simultaneous fitting by (\ref{pp}),
(\ref{pbp}) and (\ref{rhopbp}) of
$\sigma_{\mbox{\footnotesize{tot}}}^{pp}$,
$\sigma_{\mbox{\footnotesize{tot}}}^{\bar{p}p}$ and
$\rho^{\bar{p}p}$ energy dependencies.} \label{fig:tabela9}
\end{center}
\begin{center}
\begin{tabular}{l|ccccccc}
\noalign{\smallskip}\hline\noalign{\smallskip}
          &   &   &    & $\sqrt{s_{\footnotesize\mbox{min}}}$ & & & \\
Parameter & 3 & 5 & 10 & 15 & 20 & 25 & 30 \\
\hline
$k$, mbarn                 & $48 \pm 6$        & $-51 \pm 18$      & $-121 \pm 42$     & $-64 \pm 35$      & $32 \pm 14$       & $470 \pm 371$     & $400 \pm 149$ \\
$\alpha$                   & $0.025 \pm 0.006$ & $0.021 \pm 0.008$ & $0.014 \pm 0.010$ & $0.008 \pm 0.004$ & $0.022 \pm 0.009$ & $0.016 \pm 0.010$ & $0.013 \pm 0.012$ \\
$\beta \times 10^{3}$      & $3.7  \pm 0.3$    & $3.5 \pm 0.4$     & $3.1 \pm 0.5$     & $2.9 \pm 0.1$     & $3.8 \pm 0.6$     & $3.3 \pm 0.6$     & $3.1 \pm 0.7$   \\
$-\gamma$                  & $5.883 \pm 0.004$ & $5.888 \pm 0.009$ & $5.900 \pm 0.014$ & $5.908 \pm 0.002$ & $5.87 \pm 0.02$   & $5.90 \pm 0.03$   & $5.90 \pm 0.03$ \\
$\delta$, mbarn            & $51.8 \pm 0.4$    & $51.3 \pm 1.2$    & $50 \pm 2$        & $48.4 \pm 0.9$    & $56 \pm 5$        & $52 \pm 5$        & $49 \pm 5$  \\
$\bar{\alpha}$             & $0.080 \pm 0.003$ & $0.053 \pm 0.005$ & $0.038 \pm 0.010$ & $0.014 \pm 0.003$ & $0.026 \pm 0.014$ & $0.022 \pm 0.016$ & $0.019 \pm 0.013$ \\
$\bar{\beta} \times 10^{3}$& $9.1 \pm 0.3$     & $6.2 \pm 0.4$     & $4.9 \pm 0.7$     & $3.18 \pm 0.12$   & $4.0 \pm 0.9$     & $3.8 \pm 0.1$     & $3.6 \pm 0.8$  \\
$-\bar{\gamma}$            & $5.676 \pm 0.005$ & $5.763 \pm 0.011$ & $5.81 \pm 0.02$   & $5.874 \pm 0.002$ & $5.84 \pm 0.03$   & $5.84 \pm 0.04$   & $5.85 \pm 0.03$ \\
$\bar{\delta}$, mbarn      & $127.3 \pm 1.6$   & $96 \pm 3$        & $81 \pm 7$        & $63.0 \pm 1.2$    & $74 \pm 10$       & $75 \pm 14$       & $72 \pm 9$ \\
$\chi^{2}/\mbox{n.d.f.}$   & $423/306$         & $186/237$         &$113/153$          & $87.3/122$        & $62.2/87$         & $33.1/58$         & $31.4/52$  \\
\noalign{\smallskip} \hline
\end{tabular}
\caption{Parameters for global simultaneous fitting by (\ref{pp}),
(\ref{pbp}), (\ref{rhopp}) and (\ref{rhopbp}) of
$\sigma_{\mbox{\footnotesize{tot}}}^{pp}$,
$\sigma_{\mbox{\footnotesize{tot}}}^{\bar{p}p}$, $\rho^{pp}$ and
$\rho^{\bar{p}p}$ energy dependencies.}\label{fig:tabela10}
\end{center}
\begin{center}
\begin{tabular}{c|cccccccc}
\noalign{\smallskip}\hline\noalign{\smallskip}
          &                &                &                &$\sqrt{s_{\footnotesize\mbox{min}}}$ &   &    &                  \\
$\sqrt{s}$(TeV) &3        &5               &10              &15              &20             &25              &30            \\
\hline
          &                &                &                &$\sigma_{\mbox{\footnotesize{tot}}}^{pp}$ (mb)        &                & & &           \\
0.2       &51.34 $\pm$3.80&51.53$\pm$5.33 &51.55$\pm$7.12 &52.37 $\pm$1.93 &52.66$\pm$7.82 &53.69$\pm$9.60 &51.7 $\pm$10.6 \\
0.5       &59.87 $\pm$5.30&60.25$\pm$7.34 &60.32$\pm$9.69 &61.68 $\pm$2.84 &62.1 $\pm$10.3 &62.7 $\pm$12.0 &60.6 $\pm$13.7 \\
7         &91.5 $\pm$11.4 &93.3$\pm$15.8  &95.0 $\pm$20.8 &99.49 $\pm$7.10 &97.6 $\pm$20.7 &97.7 $\pm$23.1 &96.4 $\pm$27.7 \\
10        &96.3 $\pm$12.4 &98.5$\pm$17.2  &100.6$\pm$22.8 &105.73$\pm$7.89 &103.2$\pm$22.5 &103.3$\pm$25.0 &102.2$\pm$30.2 \\
14        &101.0$\pm$13.4 &103.5$\pm$18.6 &106.1$\pm$24.7 &111.84$\pm$8.68 &108.5$\pm$24.3 &108.8$\pm$27.0 &107.8$\pm$32.6 \\
30        &111.8$\pm$15.8 &115.1$\pm$22.1 &119.0$\pm$29.5 &126.5$\pm$10.7  &120.9$\pm$28.6 &121.6$\pm$31.9 &121.2$\pm$38.7 \\
40        &115.9$\pm$16.8 &119.6$\pm$23.5 &124.0$\pm$31.5 &132.3$\pm$11.5  &125.7$\pm$30.3 &126.7$\pm$33.9 &126.4$\pm$41.2 \\
50        &119.1$\pm$17.6 &123.2$\pm$24.6 &128.0$\pm$33.0 &136.9$\pm$12.1  &129.4$\pm$31.6 &130.6$\pm$35.4 &130.6$\pm$43.2 \\
100       &129.3$\pm$20.2 &134.4$\pm$28.3 &140.8$\pm$38.2 &151.7$\pm$14.3  &141.2$\pm$36.1 &143.2$\pm$40.7 &143.8$\pm$49.7 \\
200       &139.6$\pm$22.9 &145.8$\pm$32.2 &154.0$\pm$43.8 &167.2$\pm$16.8  &152.3$\pm$40.9 &156.3$\pm$46.3 &157.7$\pm$56.8 \\
500       &153.2$\pm$26.7 &161.1$\pm$37.9 &172.3$\pm$52.0 &189.0$\pm$20.4  &169.4$\pm$47.6 &174.1$\pm$54.5 &176.7$\pm$67.1 \\
\hline
          &                &                &                &$\sigma_{\mbox{\footnotesize{tot}}}^{\bar{p}p}$ (mb)        &               & & &           \\
0.546     &64.29$\pm$3.02  &62.44$\pm$5.24  &62.4 $\pm$10.7 &61.17$\pm$2.61&62.1 $\pm$15.4 &62.2 $\pm$11.9& 62.2 $\pm$14.8 \\
1.8       &71.28$\pm$3.92  &74.54$\pm$7.11  &75.7 $\pm$14.8 &75.60$\pm$3.89&76.1 $\pm$21.6 &76.8 $\pm$17.4& 77.0 $\pm$20.8 \\
\noalign{\smallskip} \hline
\end{tabular}
\caption{Predictions for $\sigma_{\mbox{\footnotesize{tot}}}$ at
various energies based on the simultaneous fits of energy
dependence of proton-proton and proton-antiproton total
cross-sections.}\label{fig:tabela11}
\end{center}
\end{table*}

\clearpage

\end{document}